\documentclass[%
 reprint,
 superscriptaddress,
 amsmath,amssymb,
 aip,
 jap,
 graphicx
]{revtex4-1}

\usepackage{graphicx}
\usepackage{dcolumn}
\usepackage{bm}
\usepackage{hyperref}
\usepackage{tabularx}
\usepackage[version=4]{mhchem}
\usepackage{ulem}
\usepackage{commath}
\usepackage{amsfonts}
\usepackage{mathtools}
\usepackage{xcolor}
\usepackage{soul}

\begin{document}



\title{Element similarity in high-dimensional materials representations}
%
%
\author{Anthony Onwuli}
\affiliation{Department of Materials, Imperial College London, London SW7 2AZ, UK}

\author{Ashish V. Hegde}
\affiliation{Department of Materials, Imperial College London, London SW7 2AZ, UK}

\author{Kevin Nguyen}
\affiliation{Department of Materials, Imperial College London, London SW7 2AZ, UK}

\author{Keith T. Butler}
\email{k.butler@qmul.ac.uk}
\affiliation{School of Engineering and Materials Science, Queen Mary University of London, London E1 4NS, UK}

\author{Aron Walsh}
\email{a.walsh@imperial.ac.uk}
\affiliation{Department of Materials, Imperial College London, London SW7 2AZ, UK}
\affiliation{Department of Physics, Ewha Womans University, Seoul 03760, Korea}

\date{\today}

\begin{abstract}
The traditional display of elements in the periodic table is convenient for the study of chemistry and physics. However, the atomic number alone is insufficient for training statistical machine learning models to describe and extract composition-structure-property relationships. Here, we assess the similarity and correlations contained within high-dimensional local and distributed representations of the chemical elements, as implemented in an open-source Python package \textsc{ElementEmbeddings}. These include element vectors of up to 200 dimensions derived from known physical properties, crystal structure analysis, natural language processing, and deep learning models. A range of distance measures are compared and a clustering of elements into familiar groups is found using dimensionality reduction techniques. The cosine similarity is used to assess the utility of these metrics for crystal structure prediction, showing that they can outperform the traditional radius ratio rules for the structural classification of AB binary solids. 
\end{abstract}

\sethlcolor{white}

\maketitle

\section{Introduction}

The periodic table offers an effective description of the elements in order of increasing atomic number. 
Its true power comes from the latent information that it contains. 
Chemists are educated to recall periodic trends in electronic configuration, atomic radius, electronegativity, accessible oxidation states, and related characteristics.
This understanding gives the ability to rapidly assess, with bias, whether a particular compound will be stable or infer what properties a molecule or material may possess without detailed computations.\cite{Pauling1948a,axonLatticeSpacingsSolid1948,phillipsBondsBandsSemiconductors1973,pettiforStructuresBinaryCompounds1986}

Significant advances have been made in the statistical description of chemical systems with the application of supervised, unsupervised and generative machine learning (ML) techniques.\cite{sanchez-lengelingInverseMolecularDesign2018,butlerMachineLearningMolecular2018,ceriottiUnsupervisedMachineLearning2019} 
A critical factor in the performance of such ML models for chemical systems is the representation of the constituent elements. 
The atomic number of an element can be augmented or replaced by a vector that may be built directly from standard data tables, trained from chemical datasets using a machine learning model, or even generated from random numbers. 
Such representations can be categorised as local (vector components with specific meaning) or distributed (vector components learned from training data).
These have been used to build powerful ML models for property prediction based on composition alone.\cite{jhaElemNetDeepLearning2018,sekoCompositionalDescriptorbasedRecommender2018,goodallPredictingMaterialsProperties2020,wangCompositionallyRestrictedAttentionbased2021}

Perhaps the simplest local representation is one-hot encoding where a binary \textit{n}-dimensional vector \textbf{v} is used to categorise the atomic number of the element, e.g. H can be represented as 
$\begin{bmatrix}1 0 0 0... \end{bmatrix}$
and He as
$\begin{bmatrix}0 1 0 0... \end{bmatrix}$.
A single component is `hot' for each element, thus providing an orthogonal and sparse description.  
A selection of other common representations from the literature is given in Table \ref{table1}.

In this study, we are interested in the latent chemical information that can be distilled from such high-dimensional element representations. 
We consider the fundamental concept of element similarity, which can be defined here as the distance or correlation between elemental vectors. 
We explore various metrics and then apply them to data-driven structure classification for the case of binary solids. The underlying tools have been combined into an open-source and modular Python package \textsc{ElementEmbeddings} to support future investigations.

\begin{table}
\caption{\label{table1} Summary of the element vector representations discussed in this work.}
\begin{ruledtabular}
\begin{tabular}{ccc}
    Name &  Dimension & Origin \\   
\hline
\hline
    Magpie\cite{wardGeneralpurposeMachineLearning2016}  & 22 &  Element properties \\
    MatScholar\cite{westonNamedEntityRecognition2019}  & 200 & Literature word embedding \\
    Mat2vec\cite{tshitoyanUnsupervisedWordEmbeddings2019}  & 200 & Literature word embedding \\
    MEGnet\cite{chenGraphNetworksUniversal2019}  & 16 & Crystal graph neutral network \\
    Oliynyk\cite{oliynykHighThroughputMachineLearningDrivenSynthesis2016}  & 44 & Element properties \\
    Random\_200  & 200 & Random numbers \\    SkipAtom\cite{antunesDistributedRepresentationsAtoms2022}  & 200 & Structure graph pooling 
\end{tabular}
\end{ruledtabular}
\end{table}

\section{Results and Discussion}

\subsection{Element representations}

We consider four vector representations of the chemical elements in the main text\hl{, but cover all seven mentioned in Table {\ref{table1}} in the final section for applications to crystal structure prediction,} with additional analysis provided as Electronic Supplementary Information (ESI). 
The aim here is not to be exhaustive but to cover a set of distinct approaches that have been developed for chemical models. 
The analysis is performed on elements 1 (H) -- 83 (Bi) as higher atomic number elements are not covered in all representation schemes. \hl{For SkipAtom, only 80 elements are considered as the noble gases Ar, He and Ne are not contained within the representation. The source of the training data for these vectors was the Materials Project, which is largely focused on inorganic crystals.}

The Magpie\cite{wardGeneralpurposeMachineLearning2016} representation is a 22-dimensional vector. 
It is a local representation \hl{where the vector components have specific meaning as they are} built from elemental properties
including atomic number, effective radii, and the row of the periodic table. 
The Mat2vec\cite{tshitoyanUnsupervisedWordEmbeddings2019} representation is a 200-dimensional vector distributed representation built from unsupervised word embeddings\cite{word2vec} of over 3 million abstracts of publications between 1922 and 2018. 
In contrast, the atomic weights from a crystal graph convolutional neural network trained to predict the formation energies of crystalline materials are used to generate the 16 dimensional 
MEGnet\cite{chenGraphNetworksUniversal2019} representation.
The Random200 representation is simply a 200-dimensional vector generated randomly for each element, \hl{employed here as a control measure.} Each vector component is generated from the standard normal distribution, $\mathcal{N}(0,1)$.

The actual vectors were collected from various sources: \hl{the Magpie, Olinyk and Mat2Vec representations were obtained as csv files from the \textsc{cbfv} repository~{\cite{cbfv}}; 
the Matscholar and MEGnet16 were obtained from the \textsc{lrcfmd/elmd} repository{\cite{eimd}}; the SkipAtom embeddings were obtained from the \textsc{lantunes/skipatom} repository;}
\textsc{numpy}\cite{2020NumPy-Array} was used to generate the Random\_200 vectors. \hl{We found that the original Oliynyk csv file had 4 columns with missing values: Miracle\_Radius\_[pm]; crystal\_radius; MB\_electronegativty; Mulliken\_EN. 
For Miracle\_Radius\_[pm], we used the mode to impute the missing values and for the other 3 columns, we used knn-imputing with the default parameters in \textsc{scikit-learn}{\cite{scikit}}. The choice of imputation was such that the overall distribution was preserved. All embedding vectors used in this work have been standardised prior to analysis.}

\subsection{Similarity measures}

The distance between two vectors depends on the choice of measure in \textit{n} dimensional space. We assess the pairwise distances between elements representations \textbf{A} and \textbf{B}.
The Minkowski distance is a metric in the normed vector space, which is a generalisation of the common distance metrics Euclidean, Manhattan and Chebyshev:
\begin{equation}
d(\textbf{A,B}) = 
(\sum_{i=1}^n |A_i - B_i|^p)^{1/p}
\end{equation}
Those three distance metrics can be derived from the Minkowski distance by appropriately choosing the exponent $p$.

For $p=2$, we obtain the Euclidean (or L2) distance which is the length of a line segment connecting \textbf{A} and \textbf{B}:
\begin{equation}
d_E(\textbf{A,B}) = 
\sqrt{
(A_1 - B_1)^2 
+ \cdots
+ (A_n - B_n)^2 }
\end{equation}
For $p=1$, the Manhattan (or L1) distance is obtained which can be defined from a sum of the absolute differences in each dimension:
\begin{equation}
d_M(\textbf{A,B}) = 
\sum_{i=1}^n |A_i - B_i|
\end{equation}
In contrast, the Chebyshev distance is obtained from the limiting case of $p \to \infty$ and takes account of the greatest one-dimensional separation across the \textit{n}-dimensional space:
\begin{equation}
d_C(\textbf{A,B}) = 
\max_i (|A_i - B_i|)
\end{equation}
Taking the example of the separation between the elements Li and K in the Magpie representation, $d_E$ = 4.09, $d_M$ = 7.87 and $d_C$ = 3.39, which shows the typical variation in absolute values.
A larger difference between Li and Bi, expected due to their placement in the periodic table, is found with $d_E$ = 9.85, $d_M$ = 37.74 and $d_C$ = 3.55.
For completeness, the Wasserstein metric (earth mover's distance), which has been adapted for materials problems,\cite{hargreavesEarthMoverDistance2020,zhang2023grouped} is also included as a function in \textsc{ElementEmbeddings} and shown in Figure S5.

\begin{figure}[]
    \centering  \includegraphics[width=8.3cm]{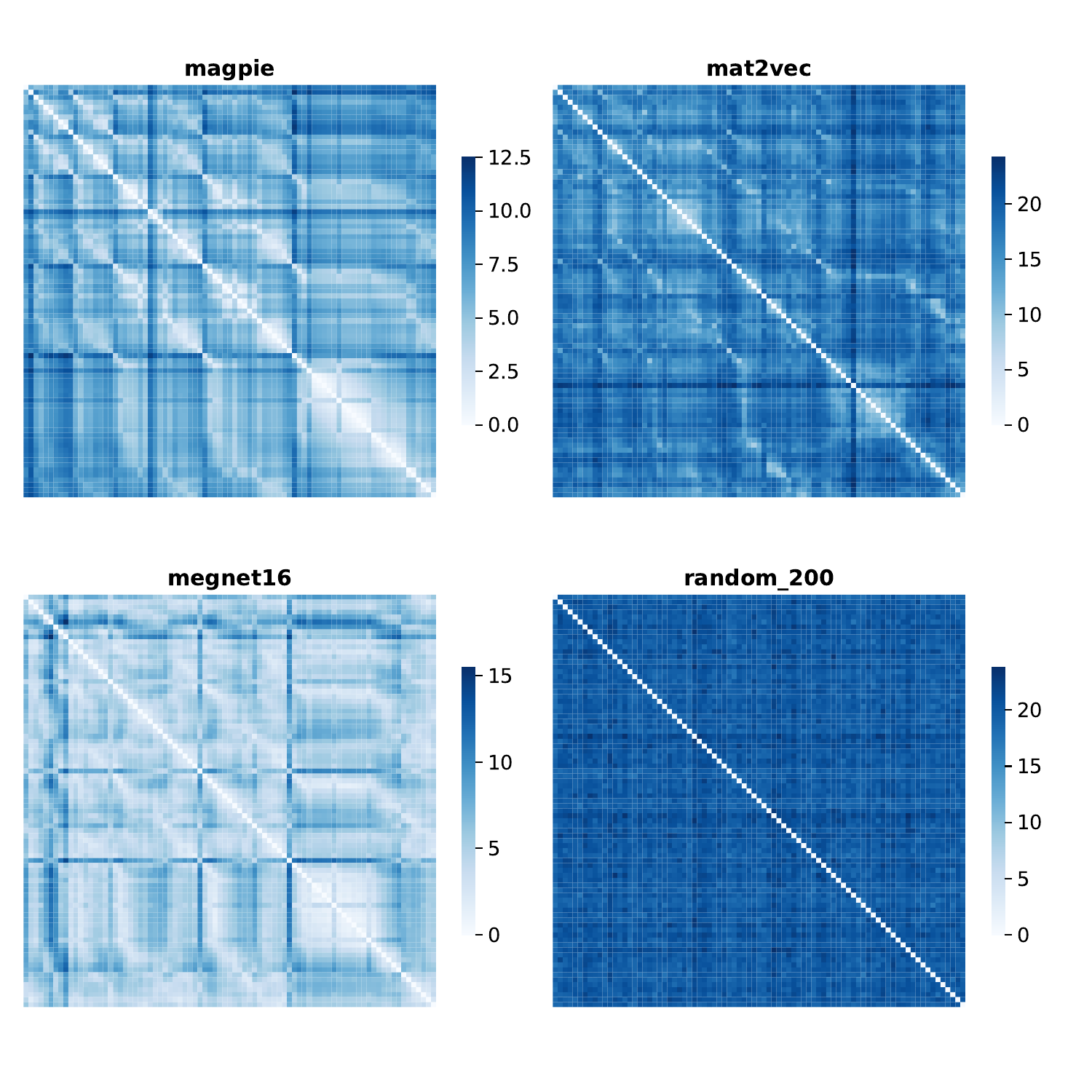}
    \caption{Map of the pairwise Euclidean distance between element vectors for four representation schemes.
    The elements are ordered in increasing atomic number along the axes from 1 (H) to 83 (Bi).
    }
    \label{f1}
\end{figure}

\begin{figure}[]
    \centering
\includegraphics[width=8.3cm]{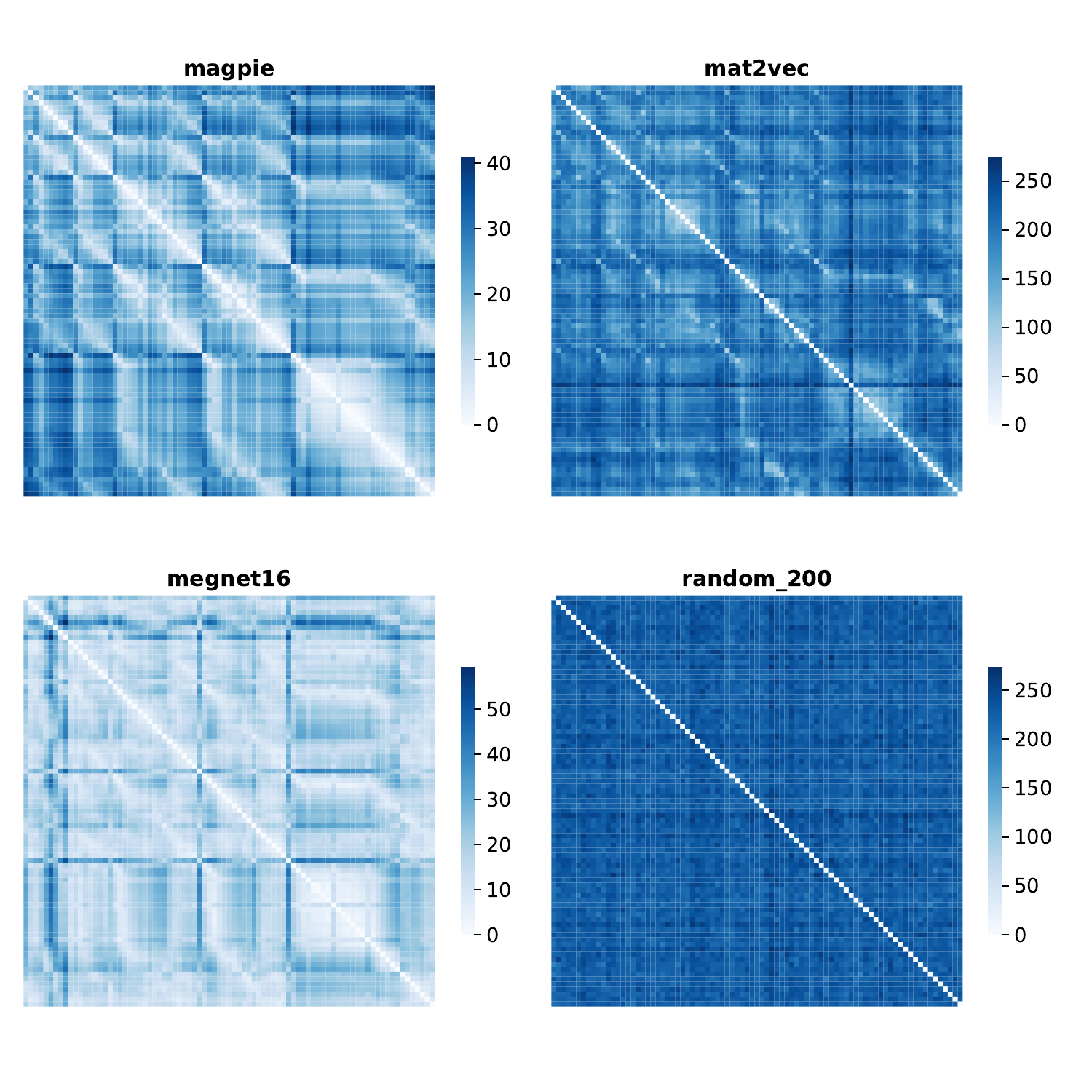}
    \caption{Map of the pairwise Manhattan distance between element vectors for four representation schemes.}
    \label{f2}
\end{figure}

Element separations are plotted for Euclidean and Manhattan distance in Figures \ref{f1} and \ref{f2}, with other measures shown in the ESI. The elements are ordered in increasing atomic number along the $x$-axis and decreasing atomic number along the $y$-axis. This cuts across the groups in the periodic table.
The leading diagonals in the distance plots are zero-valued as they correspond to $d(\textbf{A,}\textbf{A})$. The lighter blues correspond to elements whose vector representations are close to each other within the chosen metric space. 
These elements can be interpreted as similar to each other. 
Stripes are seen for the nobel gas elements, such as Kr and Xe, which are very different from the neighbouring halogens and alkali metals.
On a visual basis, the global structure of the heatmaps appears similar for the Euclidean and Manhattan distances, with the main difference being the absolute scale of the distances. Less structure is seen for the Random\_200 vectors, as expected for this control representation.

\begin{figure}[]
    \centering
 \includegraphics[width=8.3cm]{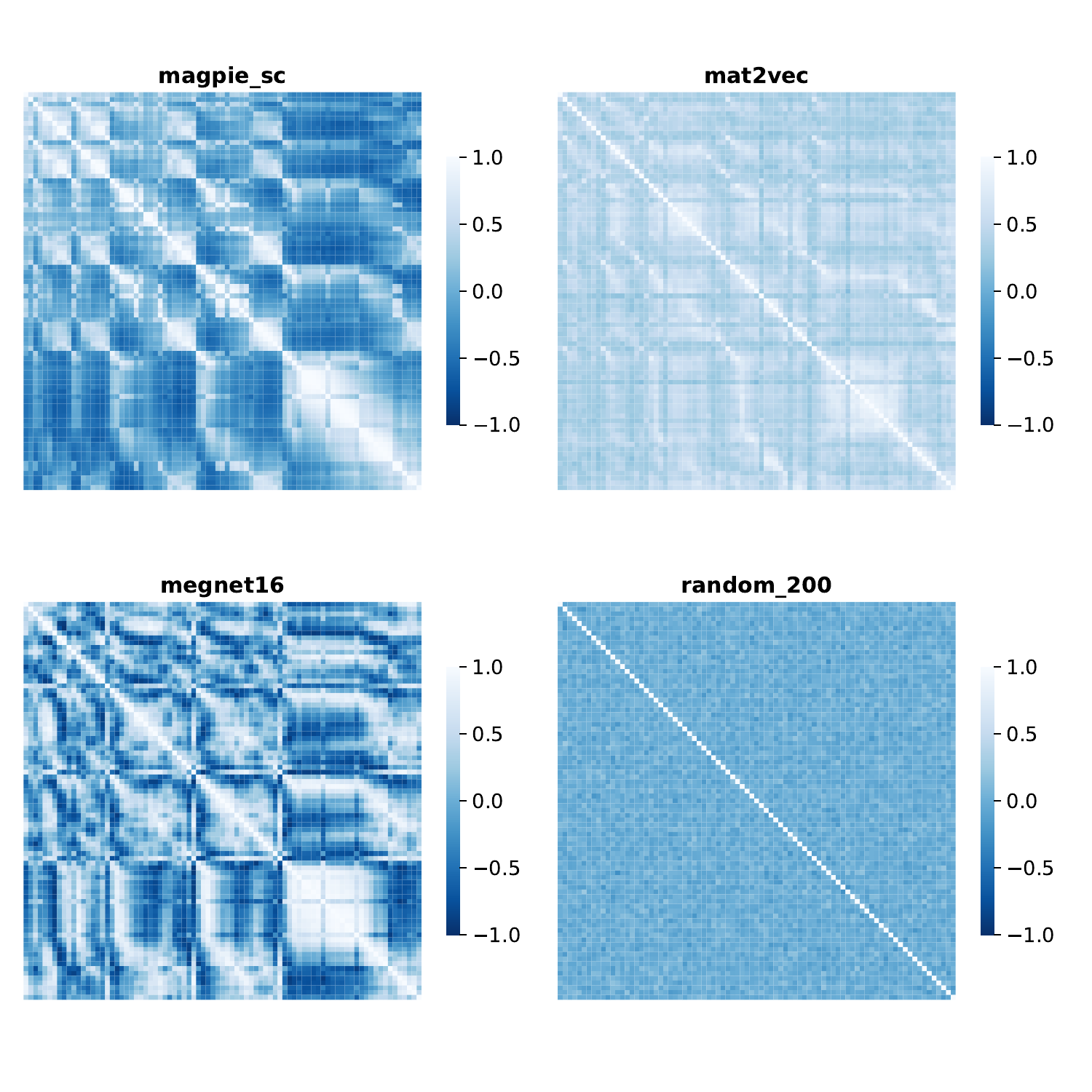}
    \caption{Map of the cosine similarity  between element vectors for four representation schemes.}
    \label{f3}
\end{figure}

Alternatively, we can 
consider the angle between vectors using the cosine similarity based on the dot product:
\begin{equation}
\cos(\theta) =  
\frac{\textbf{A} \cdot \textbf{B}}{ \norm{\textbf{A}} \norm{\textbf{B}}}
\end{equation}
For the case of Li and K, $\cos(\theta)$ = 0.738 for Magpie and -0.095 for Mat2vec.
These change to -0.603  and -0.001, respectively, for the Li and Bi pair.
The pairwise cosine similarities for the four chosen representations are shown in Figure \ref{f3}. 

\begin{figure}[]
    \centering
 \includegraphics[width=8.3cm]{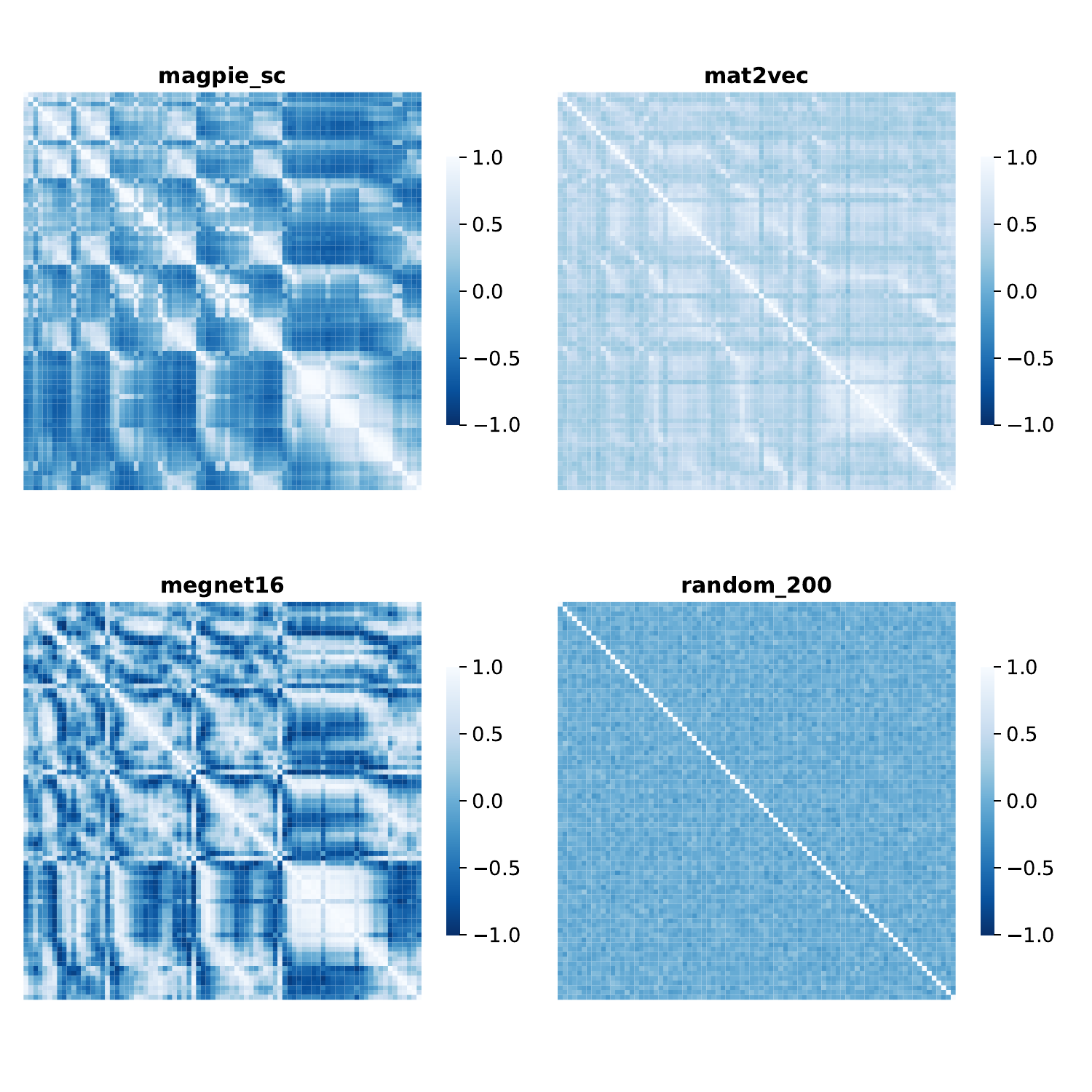}
    \caption{Map of the Pearson correlation coefficient between element vectors for four representation schemes.}
    \label{f4}
\end{figure}

The Pearson correlation coefficient provides a measure of the linear correlation:
\begin{equation}
\rho_{\textbf{A,B}} = 
\frac{cov(\textbf{A,B})}{
\sigma_{\textbf{A}}
\sigma_{\textbf{B}}}
\end{equation}
where the numerator and denominator refer to the covariance and standard deviation, respectively.
For the same case of Li and K (Bi), $\rho_{\ce{Li,K}}$ = 0.717 (-0.533) for Magpie and -0.094 (0.005) for Mat2vec. The Pearson correlation between each element is plotted in Figure \ref{f4}.

The cosine similarity and Pearson correlation are convenient metrics as both $\cos(\theta)$ and $\rho \in [-1, 1]$. The resulting heat maps are visually similar, with comparable structure to the distance metrics. 
Histograms of the values are shown in Figures S3 and S4. A skewed distribution is found in each case with the exception of Random\_200, which follows a normal distribution by construction.
We note that the cosine similarity is scale-invariant as it only depends on the angles between vectors. Some elemental representation schemes may be sensitive to bias in the training data, such as an abundance of certain metal oxides, that produce outliers in vector components. Therefore, we use cosine similarity in later sections.

\subsection{Periodic trends}

Beyond understanding the pairwise connection between elements, we can go deeper to investigate how the elements are distributed across the \textit{n} dimensions in each representation. 
For this, we use dimensionality reduction techniques based on unsupervised machine learning analysis. 
These two-dimensional plots enable intuitive interpretations of the elemental representations and aid in determining the connection to standard elemental groupings.

\begin{figure}[]
    \centering
 \includegraphics[width=8.3cm]{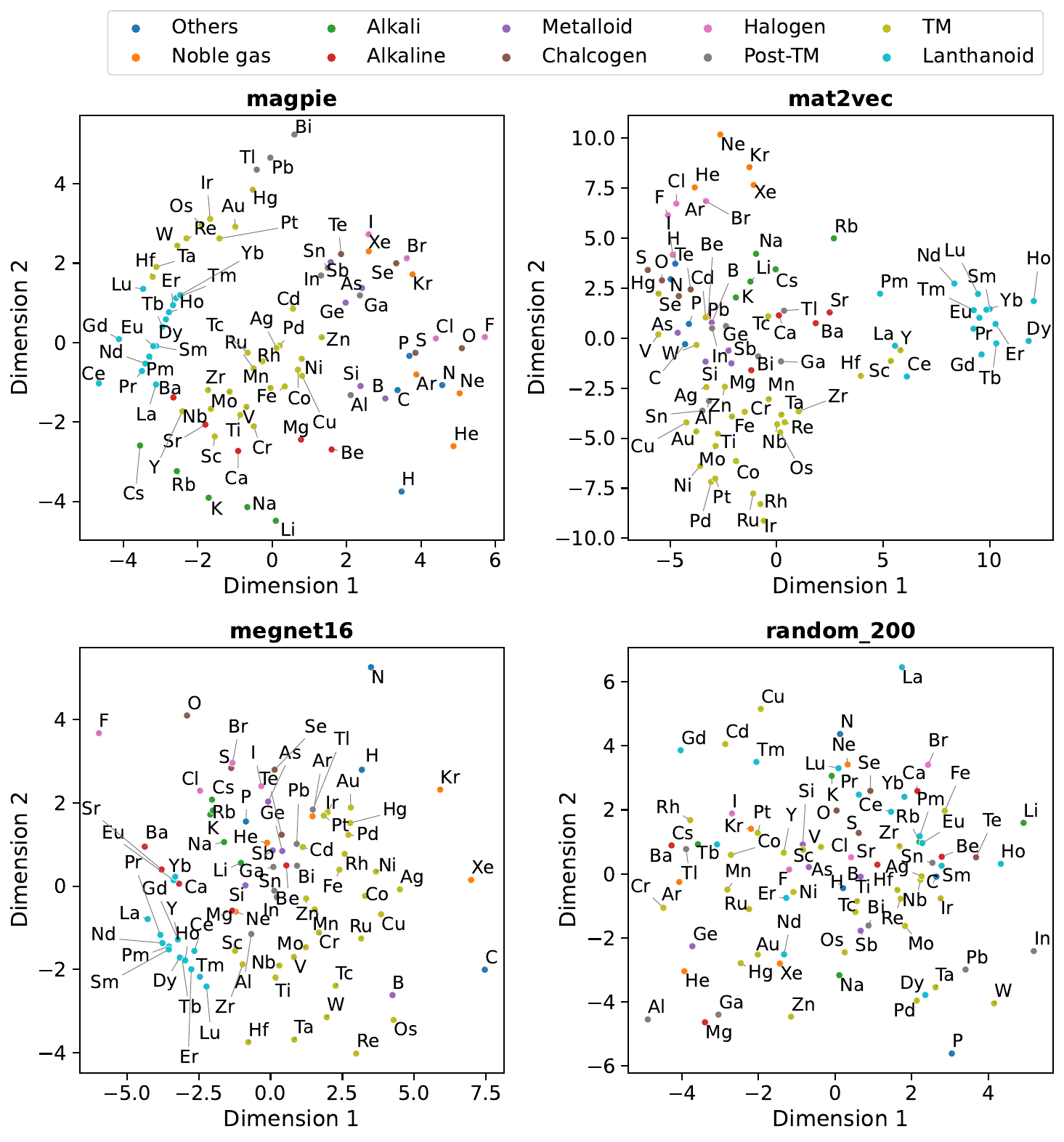}
    \caption{Two-dimensional projection of four element representations using principal component analysis.}
    \label{f5}
\end{figure}

The first method is principal component analysis (PCA). 
Here two principal component axes are defined using a linear transformation of the original features that give the greatest variance in the vector components. 
The PCA, generated using \textsc{scikit-learn}\cite{scikit}, is shown in Figure \ref{f5} with each data point coloured by the group in the periodic table.

\begin{figure}[]
    \centering
 \includegraphics[width=8.3cm]{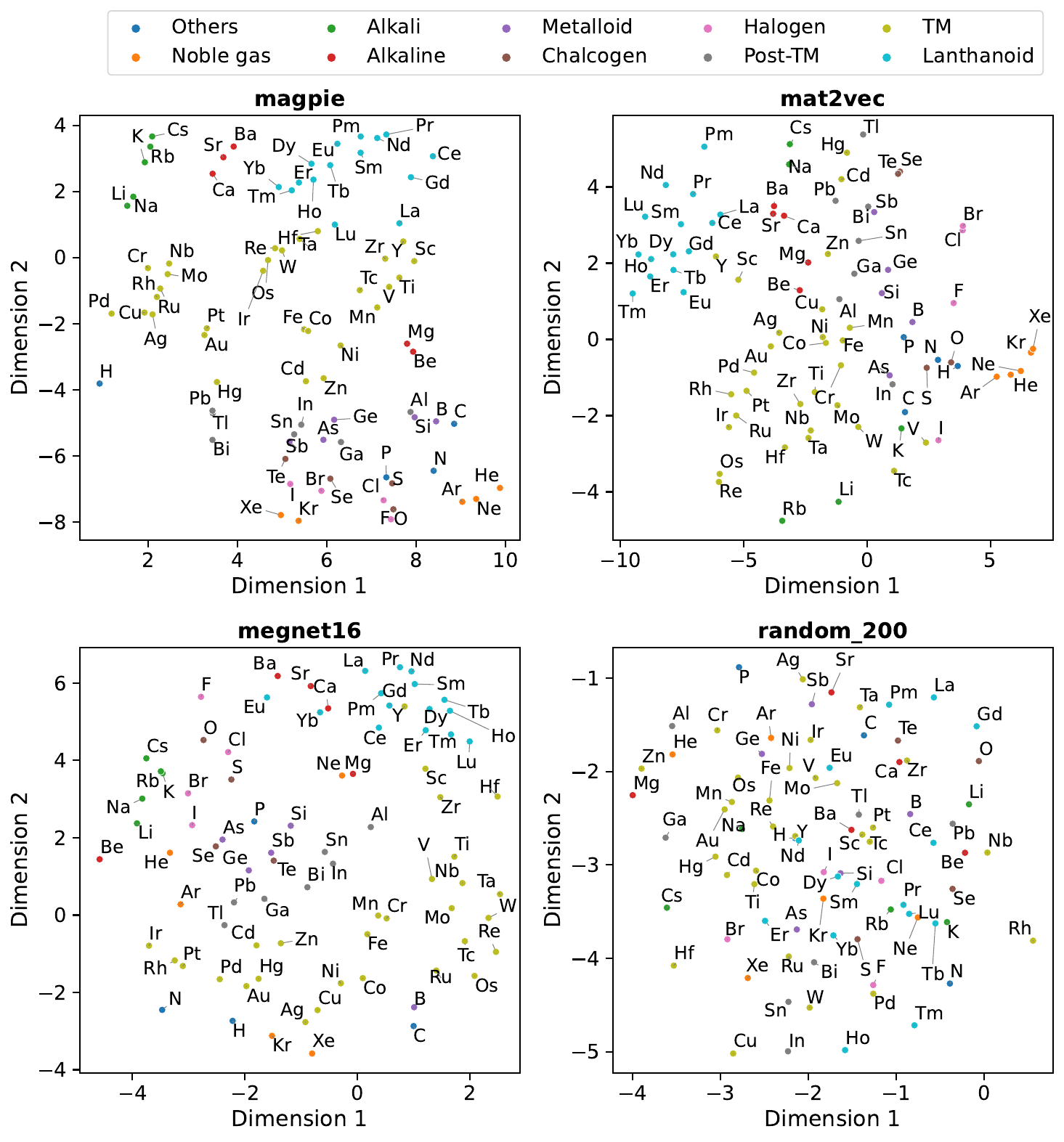}
    \caption{Two-dimensional projection of four element representations using t-SNE.}
    \label{f6}
\end{figure}

The second approach is t-distributed stochastic neighbour embedding (t-SNE). Unlike PCA, this algorithm is a nonlinear dimensionality reduction technique that can better separate data which is not linearly separable. 
Here a probability distribution is generated to represent the similarities between neighbouring points in the original high-dimensional space and a similar distribution with the same number of points is found in a lower-dimensional space. 
The t-SNE, also generated using \textsc{scikit-learn}\cite{scikit}, is shown in Figure \ref{f6} with each data point coloured by their group in the periodic table.

We observe that the element representations, with the exception of the random vectors, possess an insightful structure in the reduced dimensions, Figures \ref{f5} and \ref{f6}. The lanthanoid elements cluster together in the non-random representations independent of the choice of dimension reduction technique. In most of the representations Sr, Ba, Ca tend to group closely together, which reflects their common application in substitutional mixtures, for example in tuning ferroelectric solid-solutions. Interestingly the learned, distributed representations pick up some similarities, which are obvious to a trained chemist, but are not captured in the local Magpie representation, such as the similarity between Bi and Sb. In the Magpie representation, H tends to be considered more of an odd-one-out element, at the periphery of the distributions, whereas in the distributed representations it tends to be clustered with other elements, reflecting how it has been observed in training data from crystals such as HF and LiH. 

\subsection{Application to crystal structure prediction}\label{ssec:csp}

We have established that chemical correlations are found within the various elemental representations.
The next question is if they can be useful beyond their original purpose. 
We consider a simple classification case in crystal structure prediction, a research topic of widespread importance in computational chemistry.\cite{maddoxCrystalsFirstPrinciples1988,zungerSystematizationStableCrystal1980,pettiforStructuresBinaryCompounds1986} 

The radius ratio rules were developed to rationalise the local coordination and crystal structure preferences of ionic solids.\cite{dunitzStereochemistryIonicSolids1960}
In this model, the coordination number of a cation is determined by the balance between the electrostatic attraction (cation-anion interactions) and repulsion (anion-anion interactions).
A geometric analysis predicts that 8-fold (cubic) coordination should be obtained when the radius ratio $\rho = r_{cation}/r_{anion}$ falls in the range 0.732  -- 1.000. 
A 6-fold coordination environment is predicted for $0.414<\rho<0.732$, while 4-fold coordination is predicted for $0.225<\rho<0.414$.
For binary AB solids, these regimes are typified by the CsCl (8-fold), rocksalt (6-fold), or zinc blende/wurtzite (4-fold) structures. 
While it is accepted that there are many cases where these rules fail, especially in the lower radius ratio regime,\cite{nathanPredictionsCrystalStructure1985} they are still commonly taught in undergraduate programs due to their instructive nature. 

To assess the utility of the various element embeddings for this problem, 
we follow the structure substitution procedure proposed by Hautier et al \cite{hautierDataMinedIonic2011} and as implemented in the Python code \textsc{SMACT\hl{>=2.3}}.\cite{daviesComputationalScreeningAll2016,daviesSMACTSemiconductingMaterials2019,alex_moriarty_2021_5553202}
In this approach, the likelihood that a new chemical composition ($\textbf{X}$) will adopt the crystal structure 
of a known chemical composition ($\textbf{X}^{\prime}$) depends on the substitution probability function $p(\textbf{X},\textbf{X}^{\prime})$. 
The original pairwise substitution weights were learned from a training set of inorganic materials from the Inorganic Crystal Structure Database.\cite{bergerhoff1987crystallographic}
However, we instead use the cosine similarity between element representations, i.e. we make an assumption that the preferred crystal structure is the one that maximises $\cos(\textbf{X},\textbf{X}^{\prime})$. 

\hl{Unary substitutions are considered here, i.e. where two compositions differ by one element. This allows us to approximate the probability function to $p(\textbf{X},\textbf{X}^{\prime})=\frac{e^{\lambda}}{Z}$, where $Z$ is the partition function, and $\lambda$ is the metric for chemical similarity.  
These are the pairwise substitution weights in the original model.}\cite{hautierDataMinedIonic2011}. In the \textsc{SMACT} implementation, these can be a user-defined, pairwise metric for similarity which here is defined as $\cos(\textbf{X},\textbf{X}^{\prime})$.
A related procedure has been employed by Wang et al to predict new stable compounds\cite{glaweOptimalOneDimensional2016,wangPredictingStableCrystalline2021}, and an extension based on metric learning has been reported by Kusaba et al.\cite{kusabaCrystalStructurePrediction2022}

To obtain a set of binary AB solids that adopt one of the four structure types as their ground-state structure,
we queried the Materials Project (version: 2022.10.28)\cite{jainCommentaryMaterialsProject2013} using \textsc{pymatgen}\cite{ongPythonMaterialsGenomics2013}. \hl{The query was carried out using the parameters: formula$=*1*1$; theoretical$=$False; is\_metal$=$False. This query returned 494 binary AB solids. We chose to exclude metallic materials to focus on compositions where the bonding should be heteropolar. Some of the materials in this dataset contained polymorphs of the same composition. For example, 83 ZnS entries were returned. The data was filtered by only keeping the polymorph of a composition with the lowest energy above the convex hull as an approximation for relative stability. This filter reduced the dataset from 494 materials to 233. The query data was further filtered by matching the structures to one of the four aforementioned structure types using the structure\_matcher module in \textsc{pymatgen}{\cite{ongPythonMaterialsGenomics2013}}with the default parameters.}

Our process led to a dataset of 101 unique compounds. \hl{The final filter was to check that the remaining compounds could be assigned oxidation states, which led to a final dataset of 100 compounds.}
Taking the empirical Shannon radii\cite{shannonRevisedEffectiveIonic1976} for each ion, averaged over coordination environments, the radius ratio rules are found to correctly predict the ground-state crystal structures in 54\% of cases. \hl{This assessment was performed on 81 of the 100 compounds as Shannon radii are not available for all ions. For instance, oxygen is assigned a -1 oxidation state in AgO (mp-1079720), which has no available radius.}
The performance is lower than the 66\%  reported in a recent study of the predictive power of Pauling's rules, and using Pauling's univalent radii, to assign the coordination preferences of metals in a dataset of around 5000 metal oxides.\cite{georgeLimitedPredictivePower2020} \hl{The differences likely arise from the use of averaged Shannon radii and sensitivity to the chosen dataset.}

\begin{table}[]
\begin{ruledtabular}
\caption{Classification accuracy for the crystal structure preference of 101 binary AB solids. For comparison, the radius ratio rules, based on Shannon ionic radii, have an accuracy of 54 \%.}
\begin{tabular}{cc}
Element Embedding &  Accuracy \\
\hline
\hline
Random\_200       & \hl{58.0\%}              \\
Hautier et al.    & \hl{54.0\%}              \\
SkipAtom          & \hl{68.0\%}              \\
Oliynyk       & \hl{75.0\%}             \\
MEGNet16         & \hl{73.0\%}              \\
Magpie        & \hl{78.0\%}              \\
MatScholar        & \hl{81.0\%}              \\
Mat2vec           & \hl{80.0\%}              \\
\end{tabular}
\end{ruledtabular}
\end{table}

\begin{figure}[]
    \centering
    \includegraphics[width=7cm]{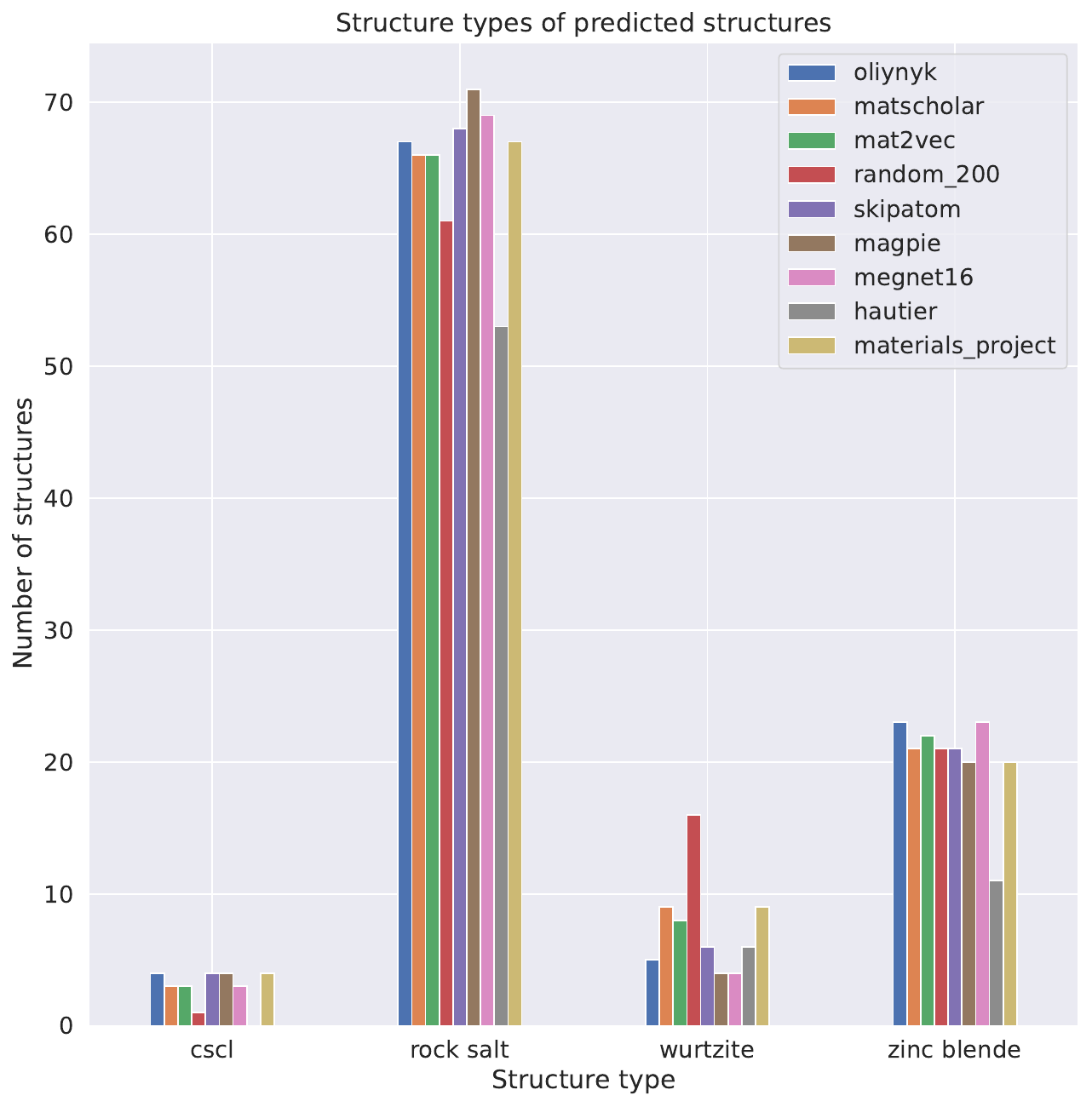}
    \caption{Performance of element representations at classifying the crystal structures of binary AB solids. The Materials Project bar refers to the ground truth label (structure at the bottom of the thermodynamic convex hull) for the 100 compositions in the dataset.}
    \label{f7}
\end{figure}

\begin{figure}[]
    \centering
    \includegraphics[width=8.5cm]{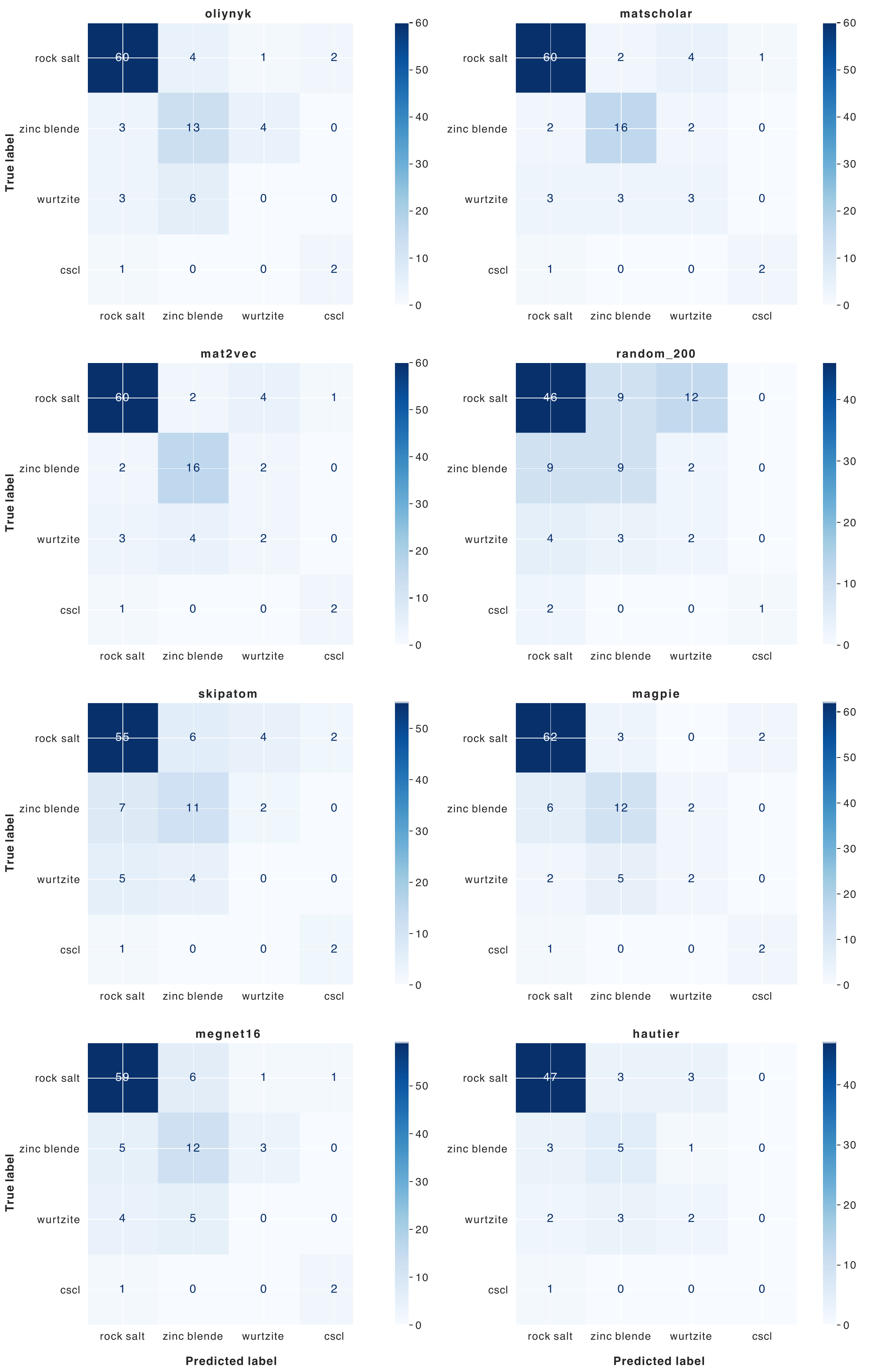}
    \caption{\hl{Confusion matrices for the classification of binary AB crystal structures for 8 element substitution (similarity) measures.}}
    \label{f8}
\end{figure}

\hl{The measure of performance defined here is classification accuracy. It is determined by the number of compositions with correctly predicted ground state structure, via the most probable substitution, over the total number of compositions in the dataset:}
\begin{equation}
\textrm{Accuracy} = \frac{\textrm{Number of correct structure types}}{\textrm{Total number of compositions}}
\end{equation}
The performance of the elemental representations ranges from\hl{ 68 to 81 \%}. Each representation performed better at this task than the previous data-mined weights of Hautier et al, with Random\_200 performing the worst. 
The classification between structure types is compared in Figure \ref{f7}, \hl{with confusion matrices shown in Figure {\ref{f8}} to further illustrate the breakdown in class predictions.} 

\hl{We find that representations derived from literature word embeddings (MatScholar and Mat2Vec) have comparable performance with their confusion matrices being almost identical. Both capture  similar correlations from the dataset of abstracts on which they were trained.
The poorer performance of the original weights from Hautier et. al{\cite{hautierDataMinedIonic2011}} can be attributed to the absence of particular oxidation states, which led to some compositions not being assigned to a structure. 
This is a limitation of species-based measures as compared to those based on the element identity alone. 
As materials databases have grown compared to a decade ago, there should be a greater diversity of compounds not included in the original training of these weights, which could extend their functionality.}

Finally, we note that while we can not exclude data leakage due to structure environments being present in the training data for some of the chosen element vectors, this particular use case has not been explicitly targeted in the training of the distributed representations. 

\section{Conclusion}

In summary, by exploring high-dimensional representations of chemical elements derived from diverse sources, we have demonstrated the potential for enhanced similarity and correlation assessments. These descriptions can complement and even outperform traditional measures, as shown in the case of crystal structure prediction and classification for binary solids. Effective chemical representations can enhance our understanding and prediction of material properties and we hope that the associated Python toolkit provided will support these developments.

~

\textit{Data availability statement:}
A repository containing the element embeddings and associated analysis code have been made available on Github (\url{https://github.com/WMD-group/ElementEmbeddings}) with a snapshot on Zenodo (DOI: 10.5281/zenodo.8101633).
The package is readily extendable to other elemental and material representations and similarity measures.

\section*{Author contributions}
The author contributions have been defined following the CRediT system. 
Conceptualisation: A.O., A.W. Investigation and Methodology: A.O., A.V.H., K.N. Software: A.O. Data curation: A.O. Supervision: A.O., K.T.B., A.W. Writing - original draft: A.O., A.W. Writing - review and editing: all authors. Resources and funding acquisition: A.W.

\acknowledgements
A.O. thanks EPSRC for a PhD studentship (EP/T51780X/1). 
We are grateful to the UK Materials and Molecular Modelling Hub for computational resources, which is partially funded by EPSRC (EP/P020194/1 and EP/T022213/1).

\bibliography{design,classics,extra}

\end{document}




\title{Electronic Supplementary Information for "Element similarity in high-dimensional materials`` representations}
%
%
\author{Anthony Onwuli}
\affiliation{Department of Materials, Imperial College London, London SW7 2AZ, UK}

\author{Ashish V. Hegde}
\affiliation{Department of Materials, Imperial College London, London SW7 2AZ, UK}

\author{Kevin Nguyen}
\affiliation{Department of Materials, Imperial College London, London SW7 2AZ, UK}

\author{Keith T. Butler}
\email{k.butler@qmul.ac.uk}
\affiliation{School of Engineering and Materials Science, Queen Mary University of London, London E1 4NS, UK}

\author{Aron Walsh}
\email{a.walsh@imperial.ac.uk}
\affiliation{Department of Materials, Imperial College London, London SW7 2AZ, UK}
\affiliation{Department of Physics, Ewha Womans University, Seoul 03760, Korea}

\date{\today}

\maketitle

\beginsupplement

\begin{figure}[]
    \centering
\includegraphics[width=14cm]{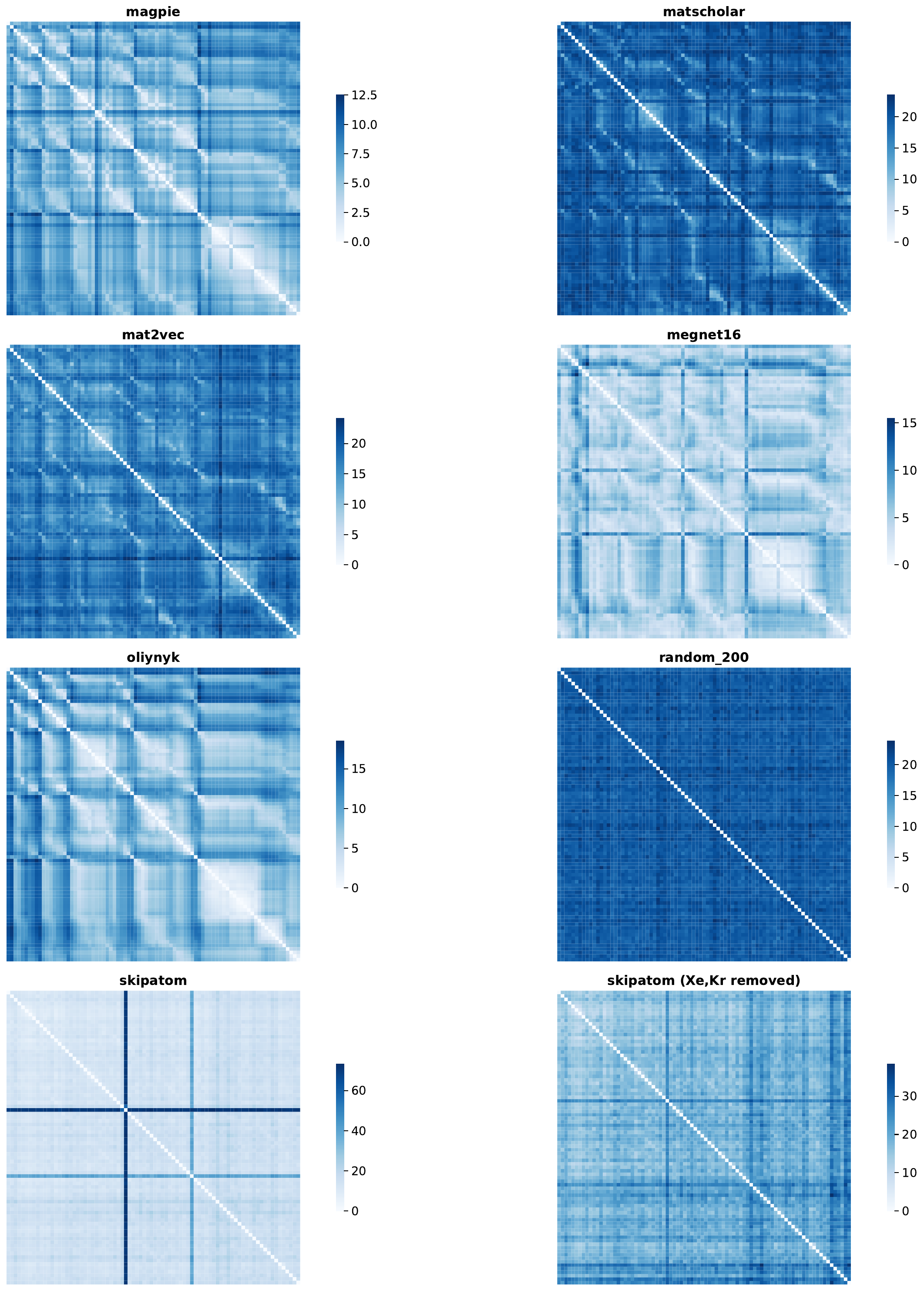}
    \caption{Map of the pairwise Euclidean distance between element vectors for seven representation schemes. The elements are ordered in increasing atomic number along the axes. The pronounced vertical and horizontal stripes in the skipatom representation, and features in others, correspond to the noble gases Kr and Xe.   
    }
    \label{s1}
\end{figure}

\begin{figure}[]
    \centering
\includegraphics[width=14cm]{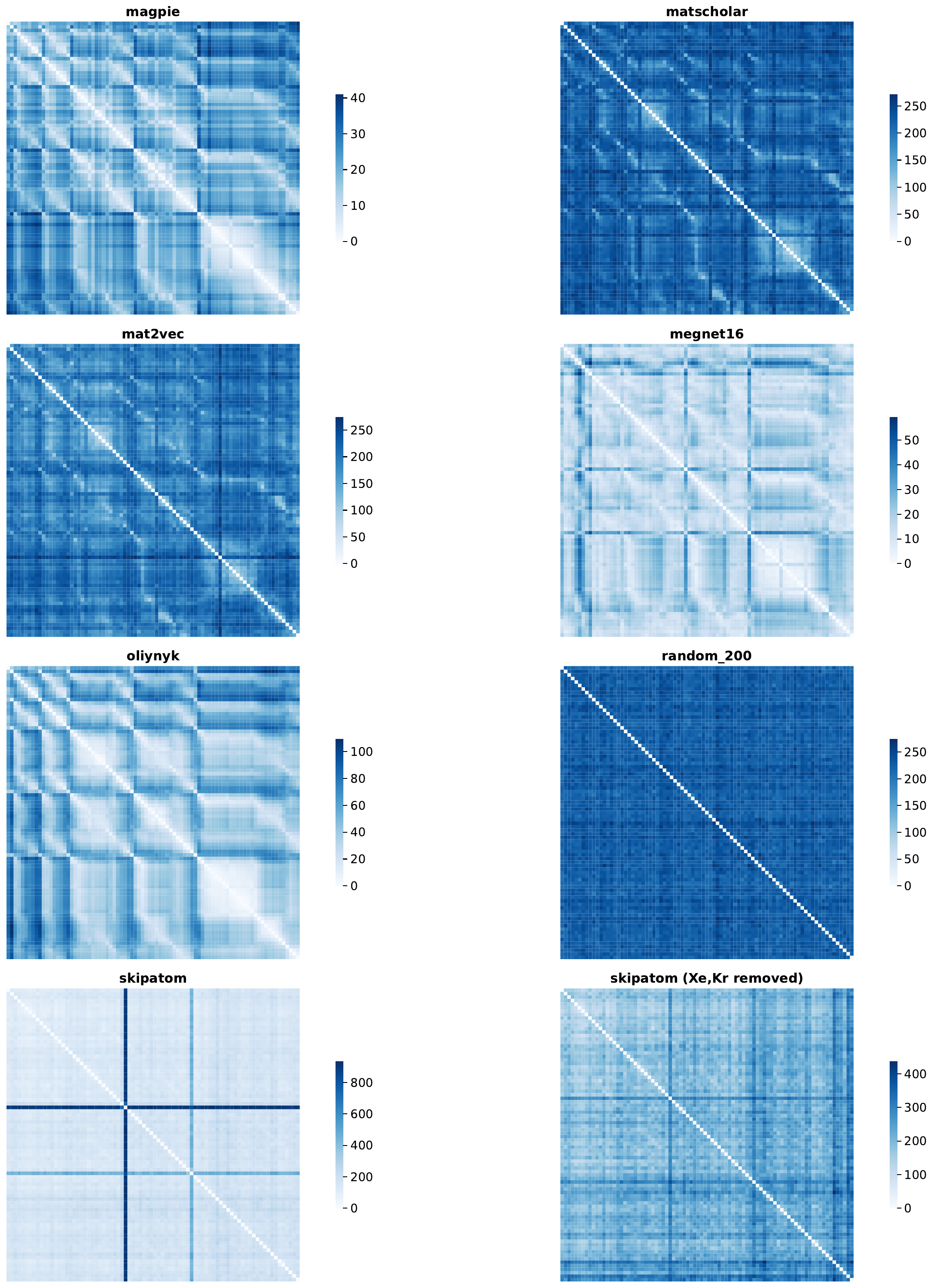}
    \caption{Map of the pairwise Manhattan distance between element vectors for seven representation schemes. The elements are ordered in increasing atomic number along the axes.}
    \label{s2}
\end{figure}

\begin{figure}[]
    \centering
\includegraphics[width=14cm]{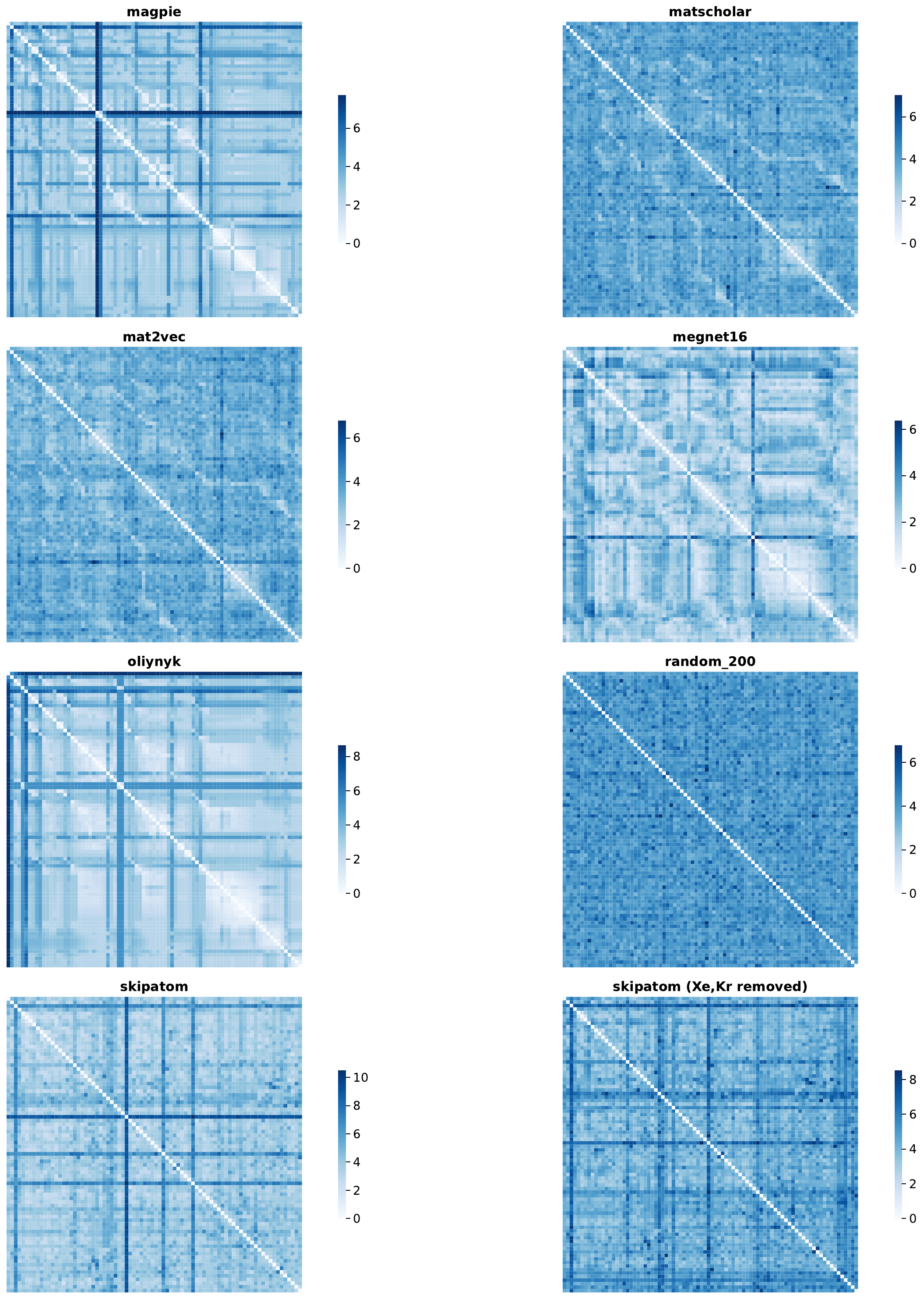}
    \caption{Map of the pairwise Chebyshev distance between element vectors for seven representation schemes. The elements are ordered in increasing atomic number along the axes.}
    \label{s3}
\end{figure}

\textbf{Supplementary Note 1: Similarity measures}
Within the main body of the text, only the Euclidean and Manhattan distances are shown for four of the similarity measures. Here, we show the distance measures currently available in \textsc{ElementEmbeddings} for seven of the representation schemes. The distance measures mentioned in the main body of the text, Euclidean, Manhattan and Chebyshev are depicted in Figures \ref{s1}, \ref{s2} and \ref{s3}, respectively.

\begin{figure}[]
    \centering
\includegraphics[width=14cm]{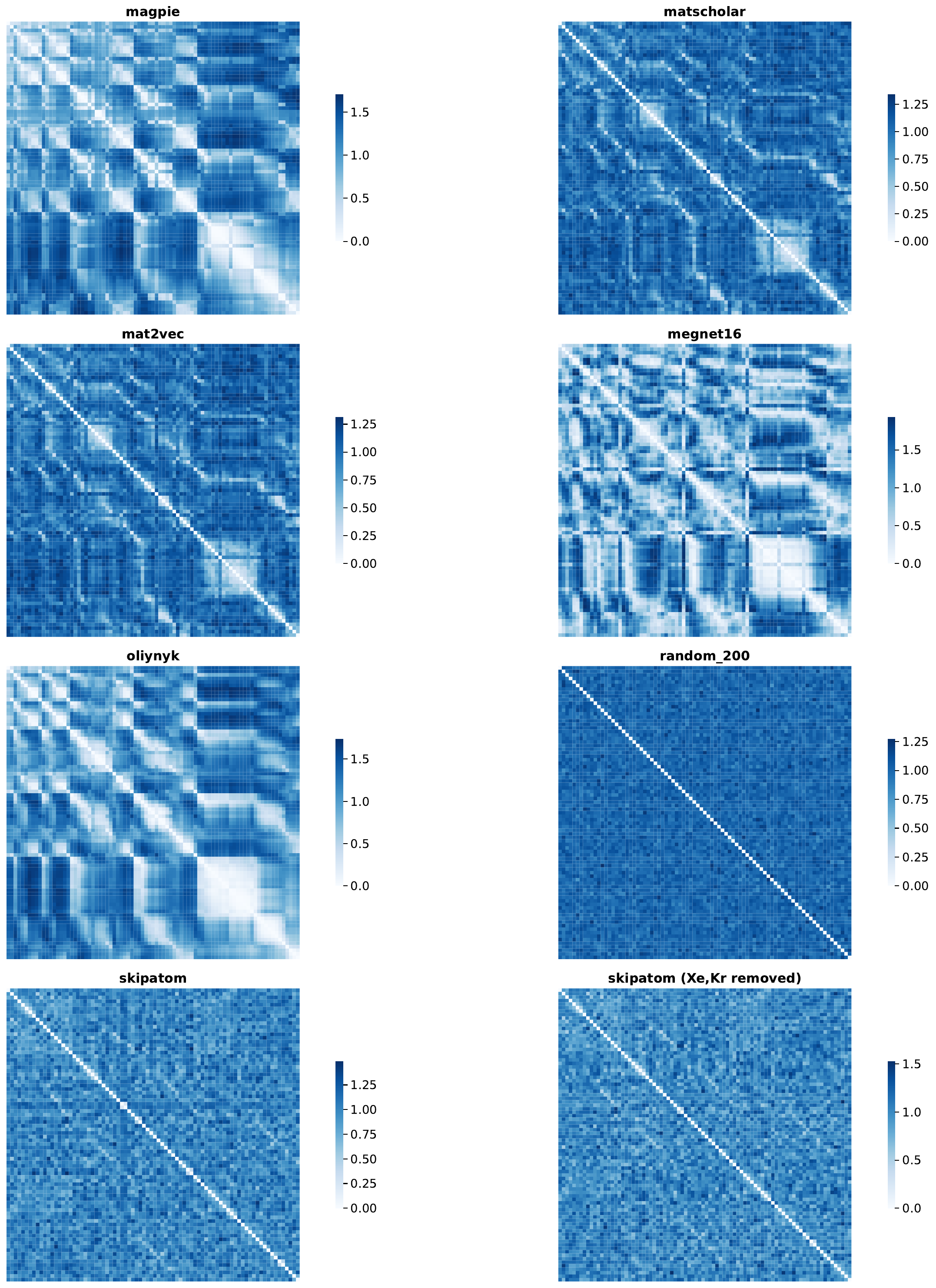}
    \caption{Map of the pairwise cosine distance between element vectors for seven representation schemes. The elements are ordered in increasing atomic number along the axes.}
    \label{s4}
\end{figure}

The cosine distance is also included as a distance measure within the package. It is the complement of the cosine similarity:
\begin{equation}
d_{cosine}= 1-cos(\theta) = 1 -  
\frac{\textbf{A} \cdot \textbf{B}}{ \norm{\textbf{A}} \norm{\textbf{B}}}
\end{equation}
The heatmaps associated with this distance measure are shown in Figure \ref{s4}.

\begin{figure}[]
    \centering
\includegraphics[width=14cm]{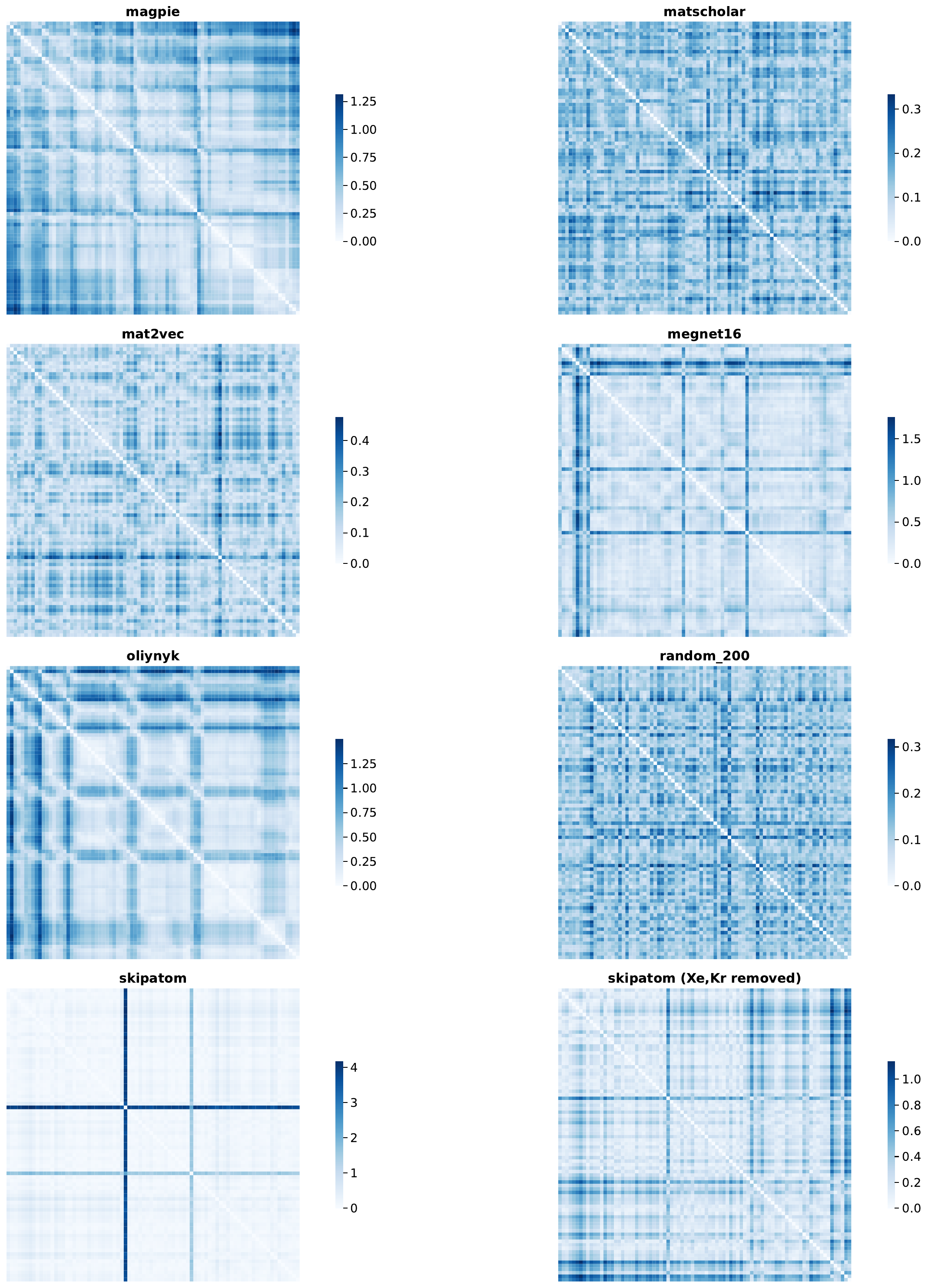}
    \caption{Map of the pairwise Wasserstein distance between element vectors for seven representation schemes. The elements are ordered in increasing atomic number along the axes.}
    \label{s5}
\end{figure}

Another metric included in the package is the Wasserstein distance which can be defined as the minimum amount of work required to transform distribution \textit{u} into \textit{v}. The first Wasserstein distance for distributions \textit{u} and \textit{v} is:

\begin{equation}
          W_1 (u, v) = \inf_{\pi \in \Gamma (u, v)} \int_{\mathbb{R} \times
        \mathbb{R}} |x-y| \mathrm{d} \pi (x, y)
\end{equation}
where $\Gamma (u, v)$ is the set of distributions on $\mathbb{R} \times \mathbb{R}$ whose marginals are \textit{u} and \textit{v} on the first and second factors, respectively. The heatmap associated with this distance measure is shown in Figure \ref{s5}.

\begin{figure}[]
    \centering
\includegraphics[width=14cm]{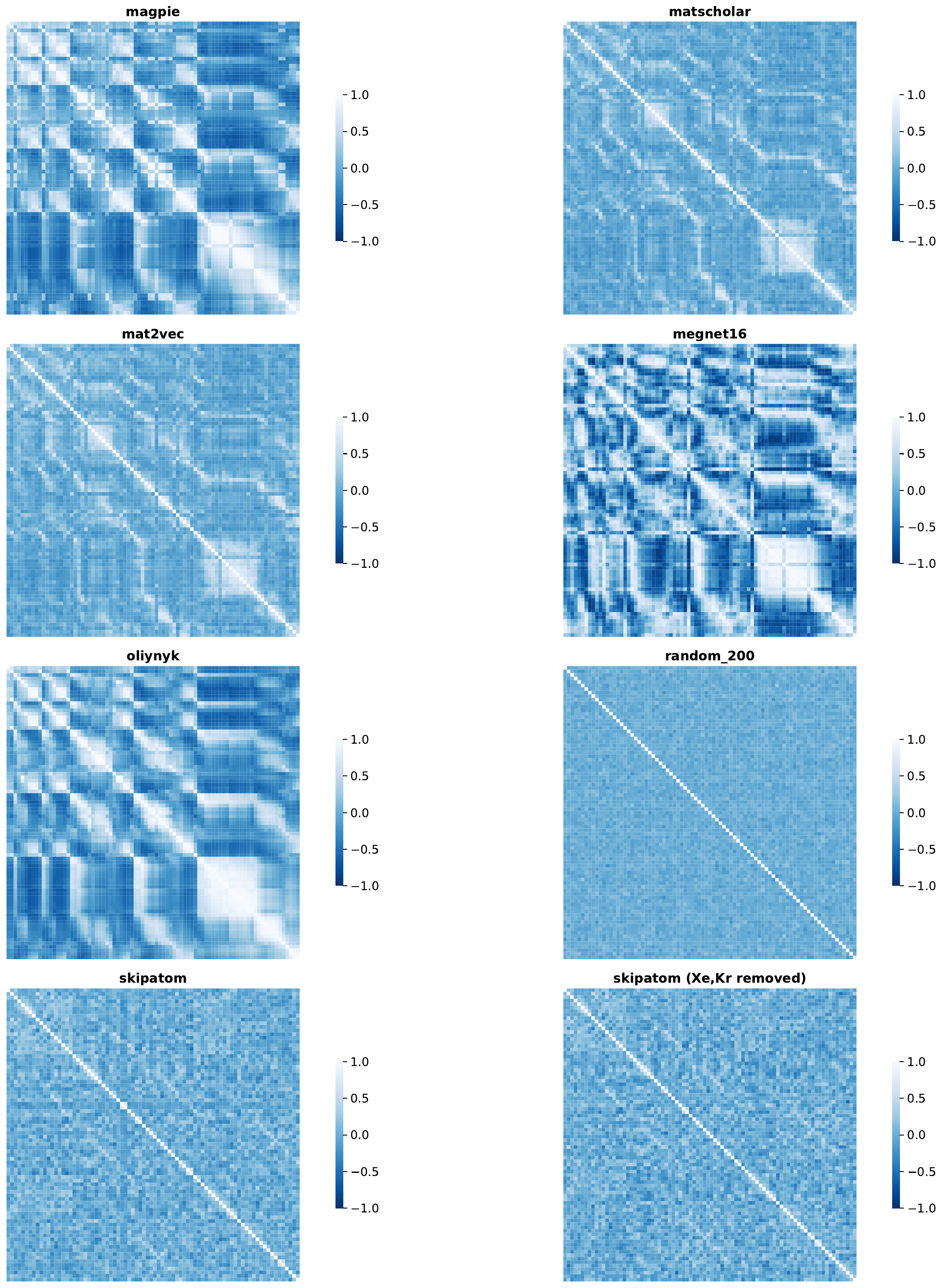}
    \caption{Map of the Pearson correlation coefficient between element vectors for seven representation schemes. The elements are ordered in increasing atomic number along the axes.}
    \label{s6}
\end{figure}

\begin{figure}[]
    \centering
\includegraphics[width=14cm]{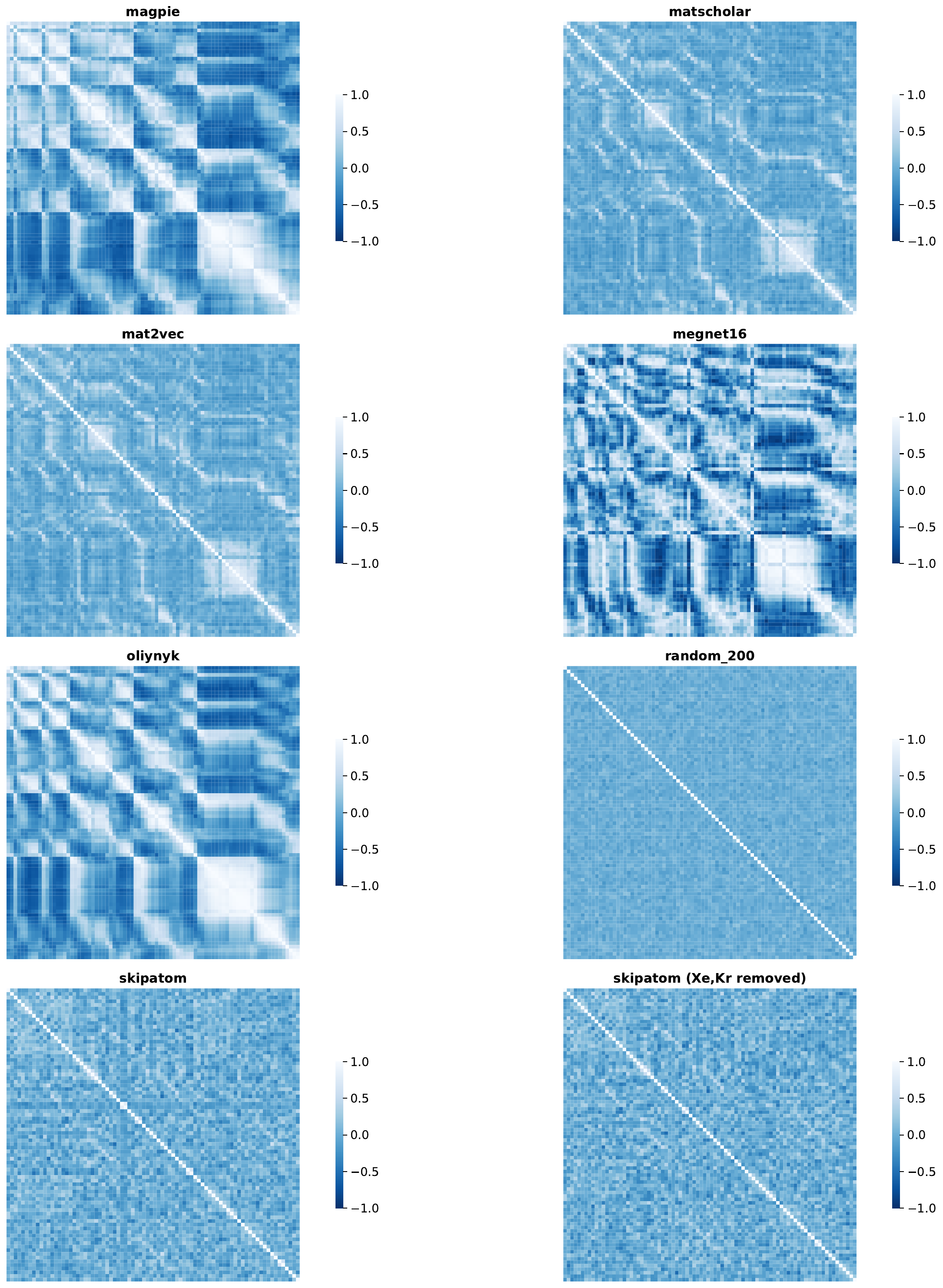}
    \caption{Map of the Spearman correlation between element vectors for seven representation schemes. The elements are ordered in increasing atomic number along the axes.}
    \label{s7}
\end{figure}

\begin{figure}[]
    \centering
\includegraphics[width=14cm]{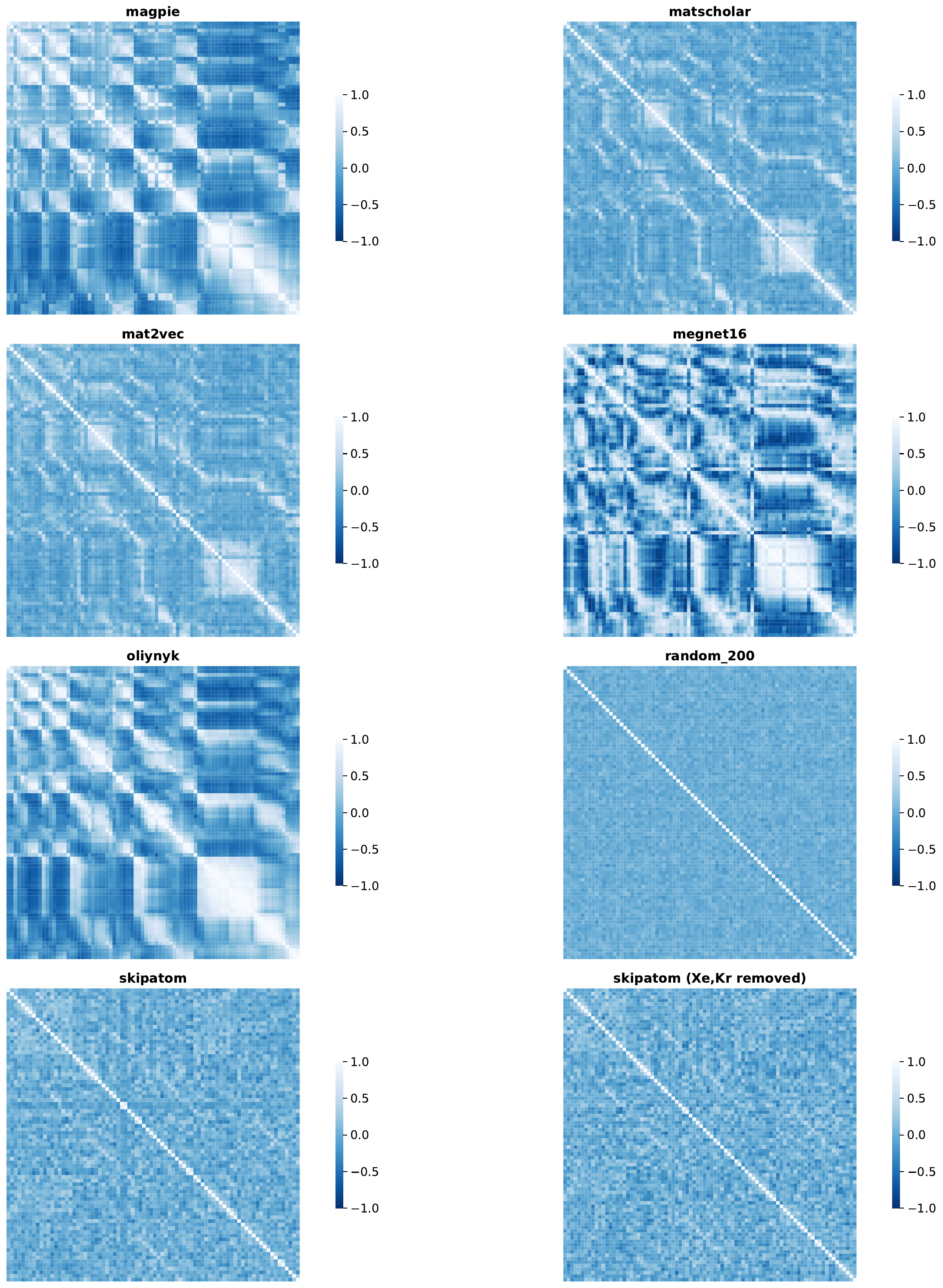}
    \caption{Map of the cosine similarity between element vectors for seven representation schemes. The elements are ordered in increasing atomic number along the axes.}
    \label{s8}
\end{figure}

The Pearson correlation and cosine similarity measures which appear in the main text are also extended to seven representation schemes. Their maps are shown in Figures \ref{s6} and \ref{s8}.
An additional correlation coefficient included in the package is the Spearman's rank correlation coefficient. This correlation measure assesses the monotonic relationship between two variables.

\begin{equation}
r_{s} = \rho_{R(\textbf{A}),R(\textbf{B})} = 
\frac{\mathrm{cov}(R(\textbf{A}),R(\textbf{B}))}{
\sigma_{R(\textbf{A})}
\sigma_{R(\textbf{B})}}
\end{equation}
Figure \ref{s7} shows the heatmaps associated with this measure.

\newpage 

\begin{figure}[]
    \centering
\includegraphics[width=14cm]{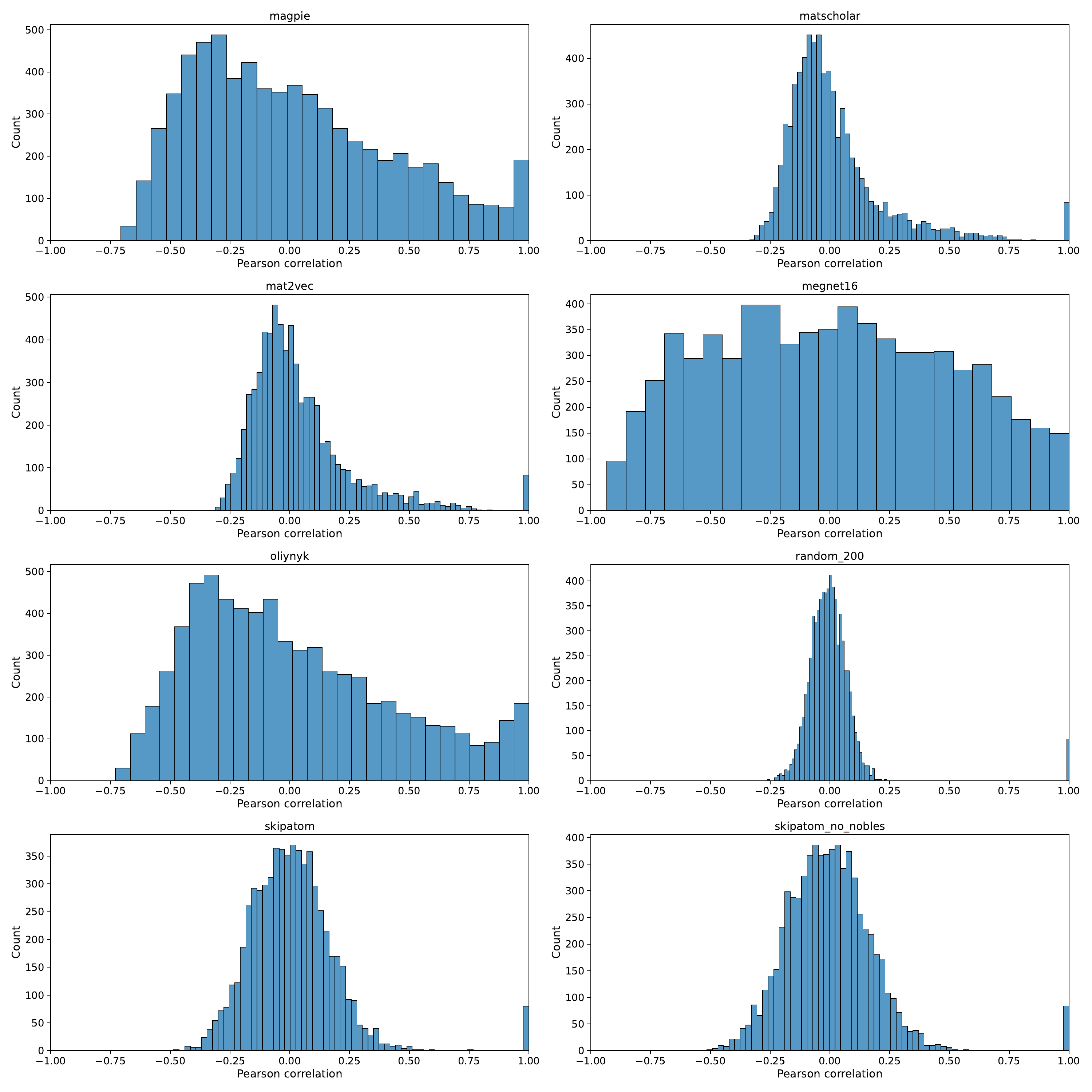}
    \caption{Distribution of the Pearson correlation between the element vectors for the four representation schemes.}
    \label{s9}
\end{figure}

\begin{figure}[]
    \centering
\includegraphics[width=14cm]{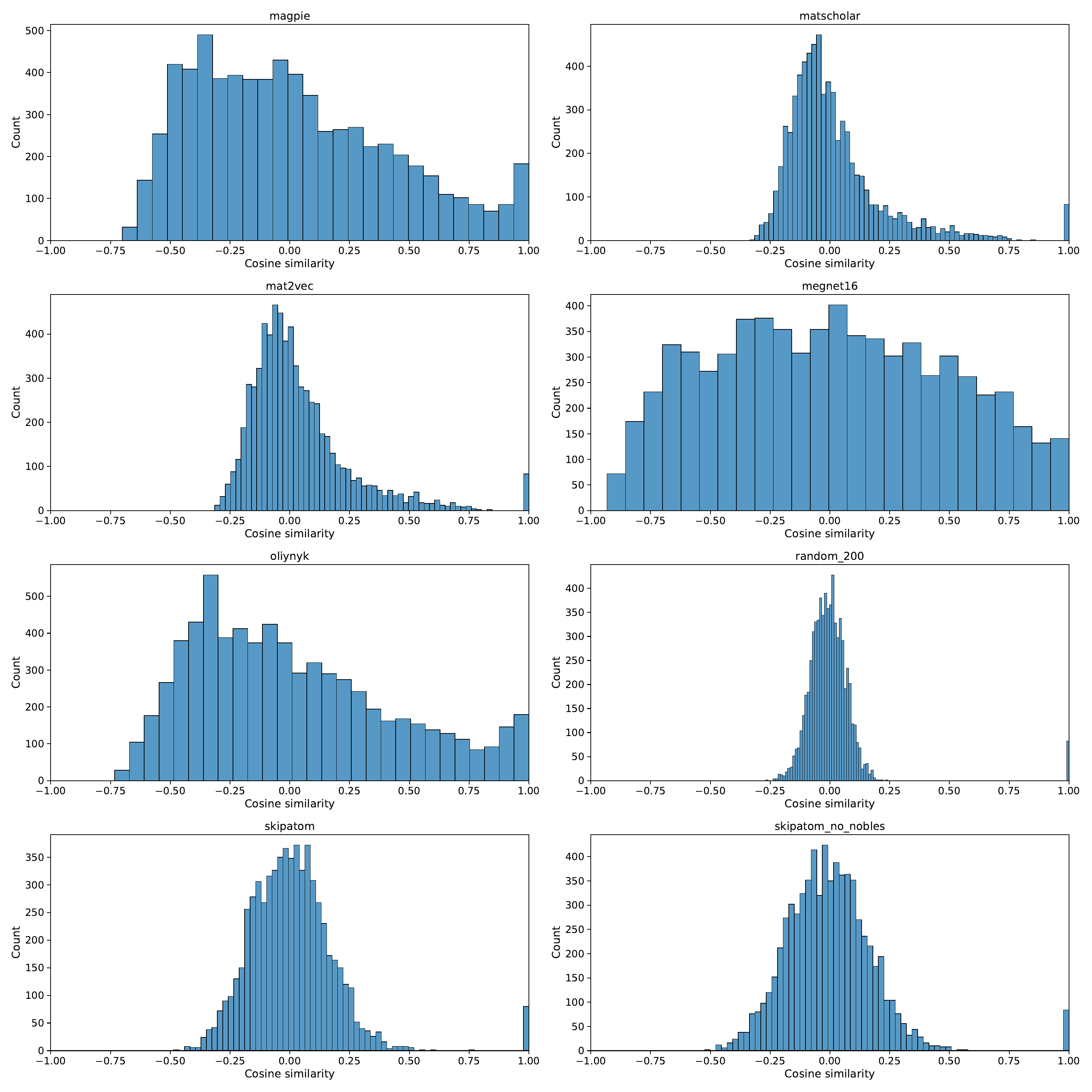}
    \caption{Distribution of the cosine similarity between the element vectors for the four representation schemes.}
    \label{s10}
\end{figure}

\textbf{Supplementary Note 2: Similarity distributions}
Figures \ref{s9} and \ref{s10} feature the distribution of the Pearson correlation coefficient and cosine similarity values between the element vectors, respectively. 

~

\begin{figure}[]
    \centering
\includegraphics[width=14cm]{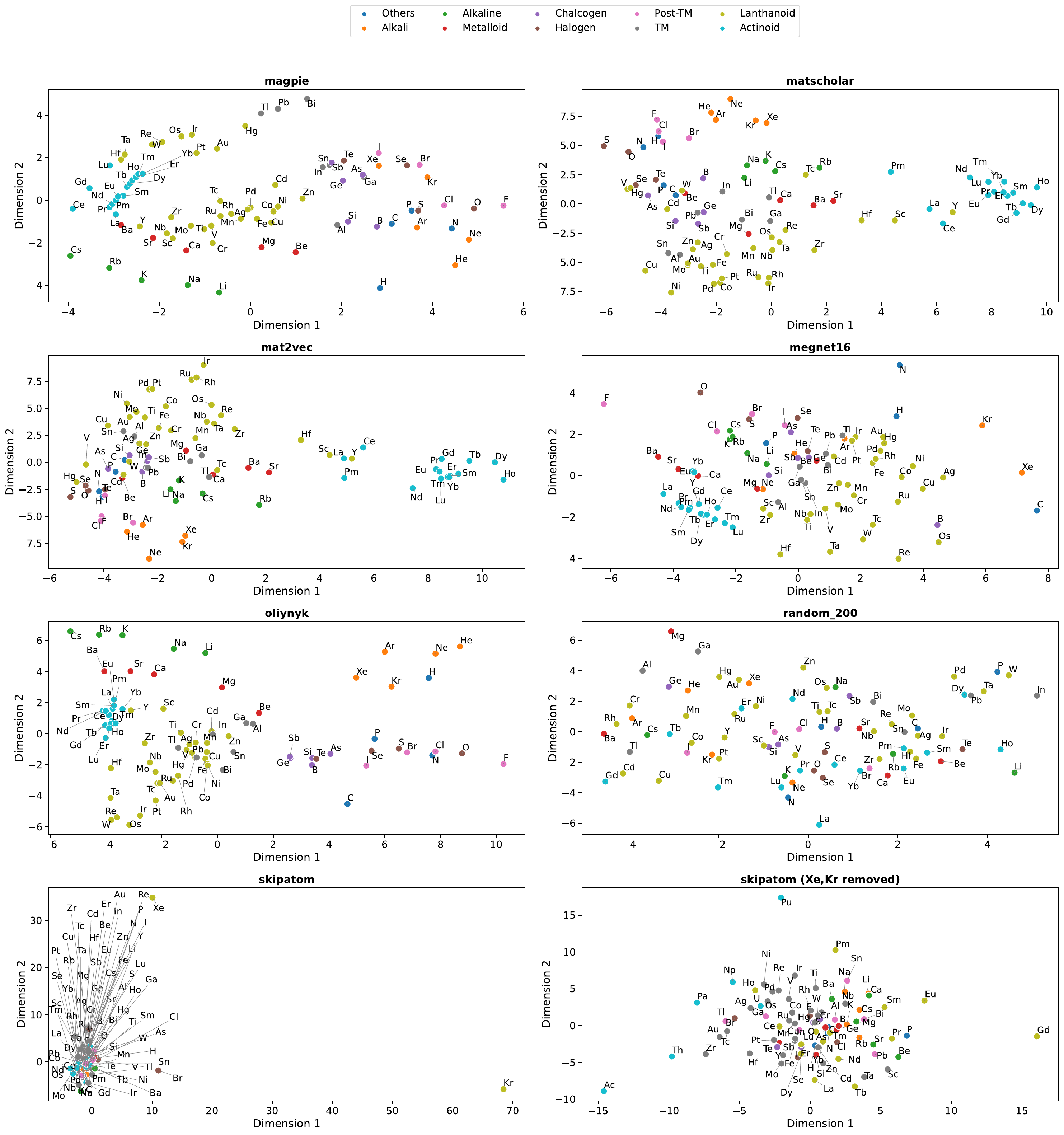}
    \caption{Two-dimensional projection of seven element representations using principal component analysis (PCA).}
    \label{s11}
\end{figure}

\begin{figure}[]
    \centering
\includegraphics[width=14cm]{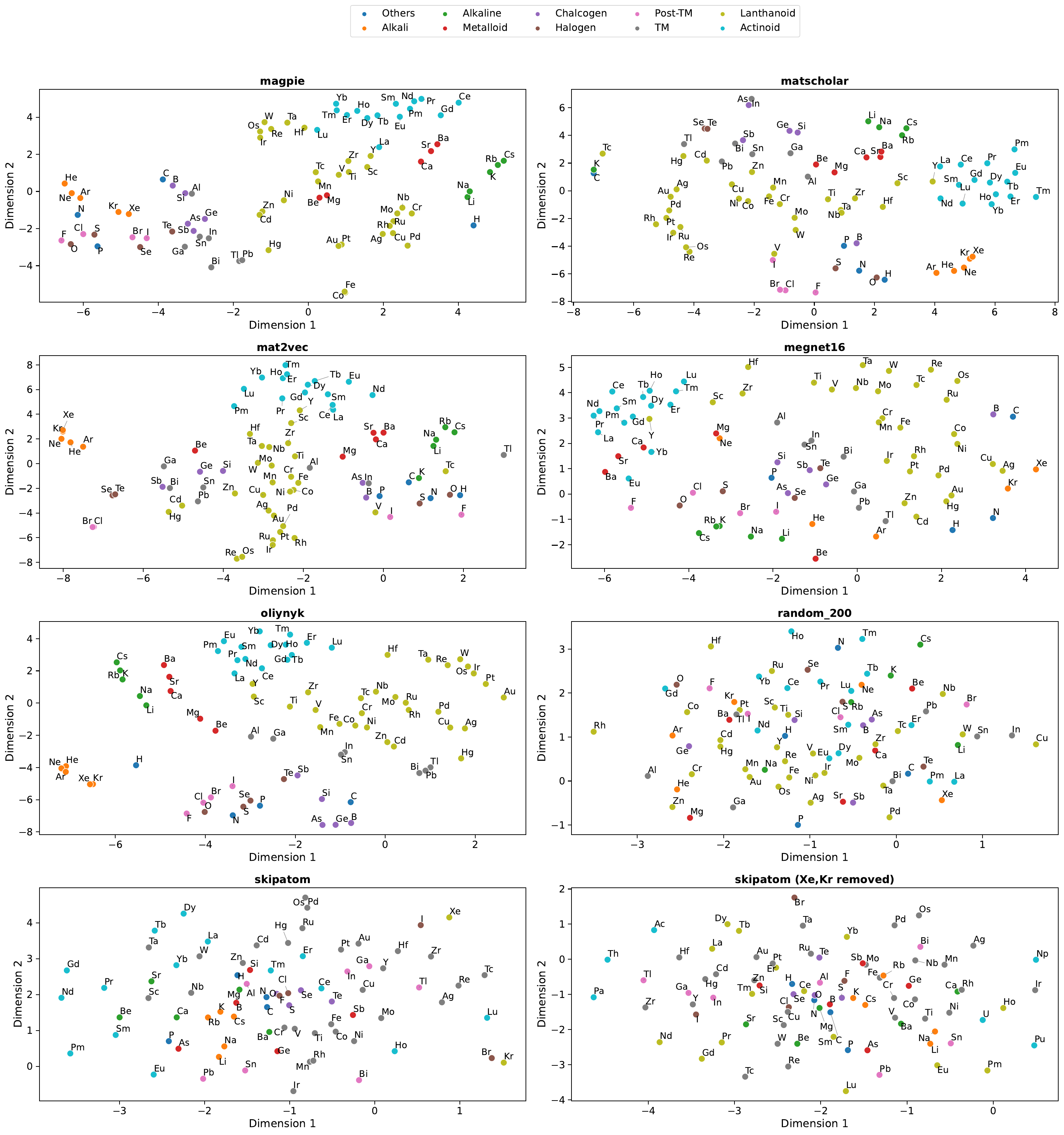}
    \caption{Two-dimensional projection of seven element representations using t-distributed stochastic neighbours (t-SNE).}
    \label{s12}
\end{figure}

\begin{figure}[]
    \centering
\includegraphics[width=14cm]{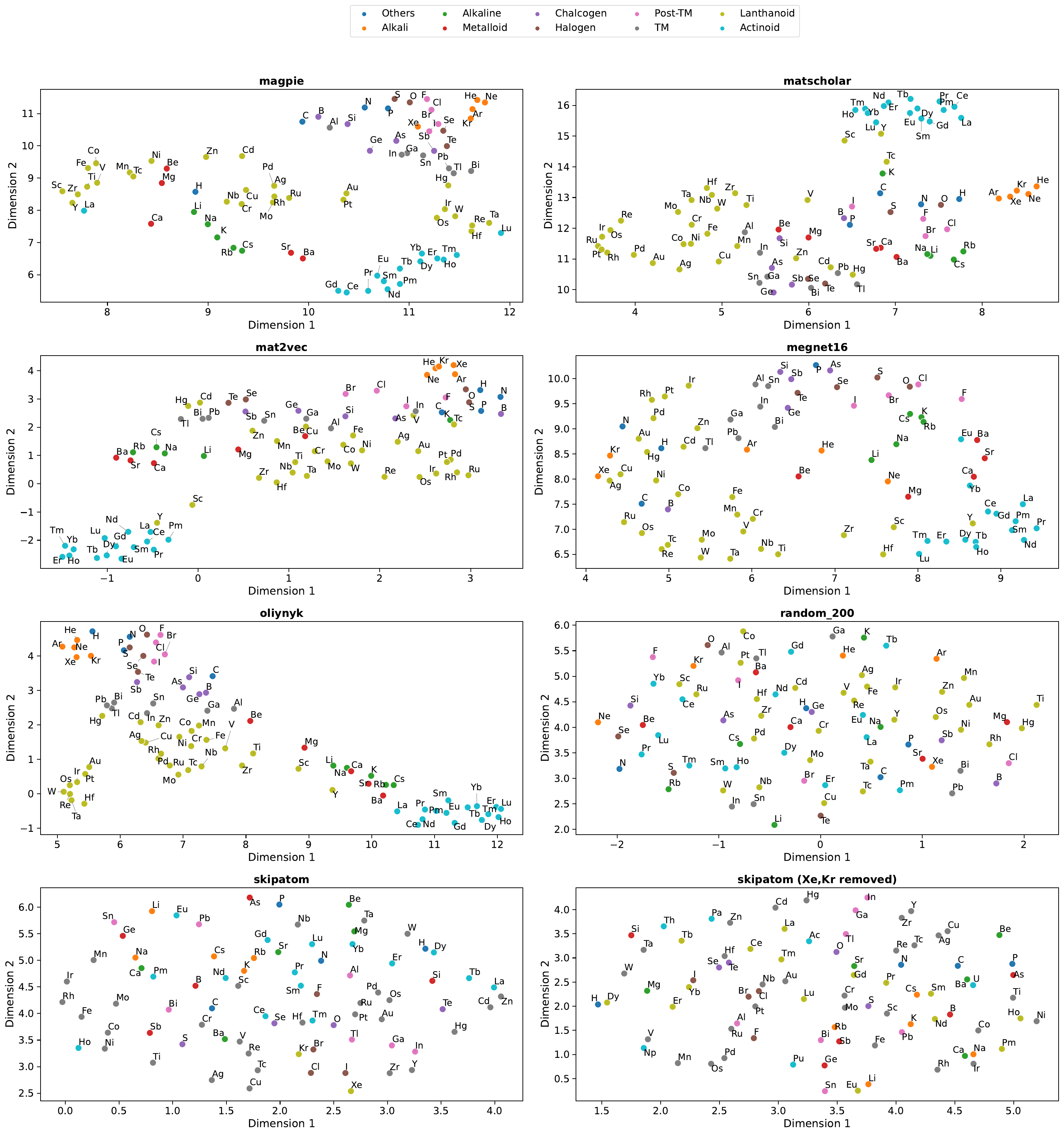}
    \caption{Two-dimensional projection of seven element representations using uniform manifold approximation and projection (UMAP).}
    \label{s13}
\end{figure}

\textbf{Supplementary Note 3: Two-dimensional projections}
The PCA and t-SNE plots which are depicted in the main body of the text are shown in Figures \ref{s11} and \ref{s12} for seven of the representations. Addtionally, a plot for UMAP is also provided in Figure \ref{s13}. For the skipatom representation in Figure \ref{s11}, the two-dimensional projection is heavily compressed due to the anomalous Kr vector.

~

\textbf{Supplementary Note 4: Crystal structure prediction }
\textsc{SMACT} was used to perform crystal structure prediction via structure substitutions of known materials from the Materials Project. Tables \ref{stab1} and \ref{stab2} show the target formulas and for each representation, the template material which the target formula was substituted in order to assign the structure to the target.


\newpage


\newpage

\pagebreak
\newpage

\begin{longtable}{lllll}
\caption{Table of the target formulas from the Materials Project and the template materials used to predict the structure of the target material. The template materials can be considered the most similar under each representation.}
\label{stab1}\\
\toprule
   target\_formula &  oliynyk\_template & matscholar\_template &  mat2vec\_template & random\_200\_template \\
\midrule
\endfirsthead
\caption[]{Table of the target formulas from the Materials Project and the template materials used to predict the structure of the target material. The template materials can be considered the most similar under each representation.} \\
\toprule
   target\_formula &  oliynyk\_template & matscholar\_template &  mat2vec\_template & random\_200\_template \\
\midrule
\endhead
\midrule
\multicolumn{5}{r}{{Continued on next page}} \\
\midrule
\endfoot

\bottomrule
\endlastfoot
  AgBr (mp-23231) &   AgCl (mp-22922) &     AgCl (mp-22922) &   AgCl (mp-22922) &      KBr (mp-23251) \\
  AgCl (mp-22922) &   AgBr (mp-23231) &     AgBr (mp-23231) &   AgBr (mp-23231) &      KCl (mp-23193) \\
   AgI (mp-22925) &   AgBr (mp-23231) &     AgBr (mp-23231) &   AgBr (mp-23231) &       KI (mp-22898) \\
 AgO (mp-1079720) &   AgCl (mp-22922) &     AgCl (mp-22922) &   AgCl (mp-22922) &      AgI (mp-22925) \\
   AlAs (mp-2172) &    GaAs (mp-2534) &      GaAs (mp-2534) &    GaAs (mp-2534) &      GaAs (mp-2534) \\
     AlN (mp-661) &     AlP (mp-1550) &        GaN (mp-804) &     AlP (mp-1550) &        GaN (mp-804) \\
    AlP (mp-1550) &      AlN (mp-661) &        AlN (mp-661) &      AlN (mp-661) &       GaP (mp-2490) \\
   AlSb (mp-2624) &     AlP (mp-1550) &      AlAs (mp-2172) &    AlAs (mp-2172) &      AlAs (mp-2172) \\
   BAs (mp-10044) &      BP (mp-1479) &        BP (mp-1479) &      BP (mp-1479) &      AlAs (mp-2172) \\
     BP (mp-1479) &    BAs (mp-10044) &      BAs (mp-10044) &    BAs (mp-10044) &       AlP (mp-1550) \\
    BaO (mp-1342) &     SrO (mp-2472) &       SrO (mp-2472) &     SrO (mp-2472) &       HgO (mp-1224) \\
    BaS (mp-1500) &     SrS (mp-1087) &       SrS (mp-1087) &     SrS (mp-1087) &        HgS (mp-634) \\
   BaSe (mp-1253) &    SrSe (mp-2758) &      SrSe (mp-2758) &    SrSe (mp-2758) &   HgSe (mp-1018722) \\
   BaTe (mp-1000) &    SrTe (mp-1958) &      SrTe (mp-1958) &    SrTe (mp-1958) &       HgTe (mp-358) \\
    BeO (mp-2542) &      BeS (mp-422) &        BeS (mp-422) &      BeS (mp-422) &      BeSe (mp-1541) \\
     BeS (mp-422) &    BeSe (mp-1541) &      BeSe (mp-1541) &    BeSe (mp-1541) &      PbS (mp-21276) \\
   BeSe (mp-1541) &      BeS (mp-422) &       BeTe (mp-252) &     BeTe (mp-252) &      PbSe (mp-2201) \\
    BeTe (mp-252) &    MgTe (mp-1039) &      BeSe (mp-1541) &    BeSe (mp-1541) &     TePb (mp-19717) \\
    CaO (mp-2605) &     SrO (mp-2472) &       SrO (mp-2472) &     SrO (mp-2472) &       YbO (mp-1216) \\
    CaS (mp-1672) &     SrS (mp-1087) &       SrS (mp-1087) &     SrS (mp-1087) &       YbS (mp-1820) \\
   CaSe (mp-1415) &    SrSe (mp-2758) &      SrSe (mp-2758) &    SrSe (mp-2758) &       YbSe (mp-286) \\
   CaTe (mp-1519) &    SrTe (mp-1958) &      SrTe (mp-1958) &    SrTe (mp-1958) &      YbTe (mp-1779) \\
     CdS (mp-672) &    CdSe (mp-2691) &        HgS (mp-634) &      HgS (mp-634) &        HgS (mp-634) \\
   CdSe (mp-2691) &      CdS (mp-672) &       CdTe (mp-406) &     CdTe (mp-406) &   HgSe (mp-1018722) \\
    CdTe (mp-406) &    ZnTe (mp-2176) &      CdSe (mp-2691) &    CdSe (mp-2691) &       HgTe (mp-358) \\
   CoO (mp-22408) &    NiO (mp-19009) &      NiO (mp-19009) &    NiO (mp-19009) &      MnO (mp-19006) \\
   CsAu (mp-2667) &   RbAu (mp-30373) &     RbAu (mp-30373) &   RbAu (mp-30373) &       CsF (mp-1784) \\
 CsBr (mp-571222) &   RbBr (mp-22867) &    CsCl (mp-573697) &  CsCl (mp-573697) &     CuBr (mp-22913) \\
 CsCl (mp-573697) &   RbCl (mp-23295) &    CsBr (mp-571222) &  CsBr (mp-571222) &     CuCl (mp-22914) \\
    CsF (mp-1784) &     RbF (mp-2064) &       RbF (mp-2064) &     RbF (mp-2064) &        NaF (mp-682) \\
 CsH (mp-1057286) &    RbH (mp-24721) &      RbH (mp-24721) &    RbH (mp-24721) &      CuH (mp-24093) \\
  CsI (mp-614603) &    RbI (mp-22903) &      RbI (mp-22903) &    RbI (mp-22903) &      NaI (mp-23268) \\
   CsTe (mp-8361) &              None &                None &              None &                None \\
  CuBr (mp-22913) &   CuCl (mp-22914) &     CuCl (mp-22914) &   CuCl (mp-22914) &      KBr (mp-23251) \\
  CuCl (mp-22914) &   CuBr (mp-22913) &     CuBr (mp-22913) &   CuBr (mp-22913) &      KCl (mp-23193) \\
   CuH (mp-24093) &   CuCl (mp-22914) &     CuCl (mp-22914) &   CuCl (mp-22914) &       KH (mp-24084) \\
    DyN (mp-1410) &      HoN (mp-883) &       TbN (mp-2117) &     TbN (mp-2117) &        GaN (mp-804) \\
   ErN (mp-19830) &     TmN (mp-1975) &        HoN (mp-883) &      HoN (mp-883) &       TmN (mp-1975) \\
   GaAs (mp-2534) &    AlAs (mp-2172) &      AlAs (mp-2172) &    AlAs (mp-2172) &      AlAs (mp-2172) \\
     GaN (mp-804) &     GaP (mp-2490) &       GaP (mp-2490) &     GaP (mp-2490) &        AlN (mp-661) \\
    GaP (mp-2490) &      GaN (mp-804) &        GaN (mp-804) &      GaN (mp-804) &       AlP (mp-1550) \\
    GeTe (mp-938) &     BeTe (mp-252) &      SnTe (mp-1883) &    SnTe (mp-1883) &       BeTe (mp-252) \\
    HgO (mp-1224) &      HgS (mp-634) &        HgS (mp-634) &      HgS (mp-634) &       MgO (mp-1265) \\
     HgS (mp-634) & HgSe (mp-1018722) &        CdS (mp-672) &      CdS (mp-672) &       MgS (mp-1315) \\
HgSe (mp-1018722) &      HgS (mp-634) &       HgTe (mp-358) &     HgTe (mp-358) &   MgSe (mp-1018040) \\
    HgTe (mp-358) &   TePb (mp-19717) &   HgSe (mp-1018722) & HgSe (mp-1018722) &      MgTe (mp-1039) \\
     HoN (mp-883) &     DyN (mp-1410) &      ErN (mp-19830) &    ErN (mp-19830) &        AlN (mp-661) \\
   InP (mp-20351) &     GaP (mp-2490) &        BP (mp-1479) &      BP (mp-1479) &       GaP (mp-2490) \\
   KBr (mp-23251) &   RbBr (mp-22867) &      KCl (mp-23193) &    KCl (mp-23193) &     CuBr (mp-22913) \\
   KCl (mp-23193) &   RbCl (mp-23295) &      KBr (mp-23251) &    KBr (mp-23251) &     CuCl (mp-22914) \\
      KF (mp-463) &     RbF (mp-2064) &      KCl (mp-23193) &    KCl (mp-23193) &        TlF (mp-720) \\
    KH (mp-24084) &    RbH (mp-24721) &         KF (mp-463) &       KF (mp-463) &      CuH (mp-24093) \\
    KI (mp-22898) &    RbI (mp-22903) &      KBr (mp-23251) &    KBr (mp-23251) &      AgI (mp-22925) \\
  LiBr (mp-23259) &   NaBr (mp-22916) &     LiCl (mp-22905) &   LiCl (mp-22905) &    TlBr (mp-568560) \\
  LiCl (mp-22905) &   NaCl (mp-22862) &     LiBr (mp-23259) &   LiBr (mp-23259) &    TlCl (mp-569639) \\
    LiF (mp-1138) &      NaF (mp-682) &        NaF (mp-682) &      NaF (mp-682) &        TlF (mp-720) \\
   LiH (mp-23703) &    NaH (mp-23870) &      NaH (mp-23870) &    NaH (mp-23870) &      RbH (mp-24721) \\
  LiI (mp-570935) &    NaI (mp-23268) &      NaI (mp-23268) &    NaI (mp-23268) &     TlI (mp-571102) \\
    LuN (mp-1102) &    ErN (mp-19830) &        HoN (mp-883) &      HoN (mp-883) &        YN (mp-2114) \\
    MgO (mp-1265) &     MgS (mp-1315) &       CaO (mp-2605) &     CaO (mp-2605) &       ZnO (mp-2133) \\
    MgS (mp-1315) & MgSe (mp-1018040) &       CaS (mp-1672) &     CaS (mp-1672) &        HgS (mp-634) \\
MgSe (mp-1018040) &     MgS (mp-1315) &      MgTe (mp-1039) &    MgTe (mp-1039) &      ZnSe (mp-1190) \\
   MgTe (mp-1039) &    CaTe (mp-1519) &   MgSe (mp-1018040) & MgSe (mp-1018040) &      ZnTe (mp-2176) \\
   MnO (mp-19006) &    MnSe (mp-2293) &       ZnO (mp-2133) &     ZnO (mp-2133) &      MnSe (mp-2293) \\
   MnSe (mp-2293) &    MnO (mp-19006) &      ZnSe (mp-1190) &    ZnSe (mp-1190) &      CdSe (mp-2691) \\
  NaBr (mp-22916) &    KBr (mp-23251) &     NaCl (mp-22862) &   NaCl (mp-22862) &    CsBr (mp-571222) \\
  NaCl (mp-22862) &    KCl (mp-23193) &     NaBr (mp-22916) &   NaBr (mp-22916) &    CsCl (mp-573697) \\
     NaF (mp-682) &       KF (mp-463) &       LiF (mp-1138) &     LiF (mp-1138) &       CsF (mp-1784) \\
   NaH (mp-23870) &     KH (mp-24084) &      LiH (mp-23703) &    LiH (mp-23703) &    CsH (mp-1057286) \\
   NaI (mp-23268) &     KI (mp-22898) &     LiI (mp-570935) &   LiI (mp-570935) &     CsI (mp-614603) \\
   NiO (mp-19009) &    CoO (mp-22408) &      CoO (mp-22408) &    CoO (mp-22408) &       SrO (mp-2472) \\
   PbS (mp-21276) &    PbSe (mp-2201) &        CdS (mp-672) &      CdS (mp-672) &        HgS (mp-634) \\
   PbSe (mp-2201) &    PbS (mp-21276) &     TePb (mp-19717) &   TePb (mp-19717) &   HgSe (mp-1018722) \\
  RbAu (mp-30373) &    CsAu (mp-2667) &      CsAu (mp-2667) &    CsAu (mp-2667) &       RbF (mp-2064) \\
  RbBr (mp-22867) &  CsBr (mp-571222) &     RbCl (mp-23295) &   RbCl (mp-23295) &     LiBr (mp-23259) \\
  RbCl (mp-23295) &  CsCl (mp-573697) &     RbBr (mp-22867) &   RbBr (mp-22867) &     LiCl (mp-22905) \\
    RbF (mp-2064) &     CsF (mp-1784) &       CsF (mp-1784) &     CsF (mp-1784) &       LiF (mp-1138) \\
   RbH (mp-24721) &  CsH (mp-1057286) &    CsH (mp-1057286) &  CsH (mp-1057286) &      LiH (mp-23703) \\
   RbI (mp-22903) &   CsI (mp-614603) &     CsI (mp-614603) &   CsI (mp-614603) &     LiI (mp-570935) \\
   SnTe (mp-1883) &     CdTe (mp-406) &      ZnTe (mp-2176) &    ZnTe (mp-2176) &       BeTe (mp-252) \\
    SrO (mp-2472) &     BaO (mp-1342) &       CaO (mp-2605) &     BaO (mp-1342) &       MgO (mp-1265) \\
    SrS (mp-1087) &     BaS (mp-1500) &       CaS (mp-1672) &     BaS (mp-1500) &       MgS (mp-1315) \\
   SrSe (mp-2758) &    BaSe (mp-1253) &      CaSe (mp-1415) &    BaSe (mp-1253) &   MgSe (mp-1018040) \\
   SrTe (mp-1958) &    BaTe (mp-1000) &      CaTe (mp-1519) &    BaTe (mp-1000) &      MgTe (mp-1039) \\
    TbN (mp-2117) &     DyN (mp-1410) &       DyN (mp-1410) &     DyN (mp-1410) &        AlN (mp-661) \\
  TePb (mp-19717) &     HgTe (mp-358) &      PbSe (mp-2201) &    PbSe (mp-2201) &       HgTe (mp-358) \\
 TlBr (mp-568560) &  TlCl (mp-569639) &    TlCl (mp-569639) &  TlCl (mp-569639) &     LiBr (mp-23259) \\
 TlCl (mp-569639) &  TlBr (mp-568560) &    TlBr (mp-568560) &  TlBr (mp-568560) &     LiCl (mp-22905) \\
     TlF (mp-720) &  TlCl (mp-569639) &    TlCl (mp-569639) &  TlCl (mp-569639) &       LiF (mp-1138) \\
  TlI (mp-571102) &  TlBr (mp-568560) &     CsI (mp-614603) &   CsI (mp-614603) &     LiI (mp-570935) \\
    TmN (mp-1975) &      HoN (mp-883) &        HoN (mp-883) &      HoN (mp-883) &      ErN (mp-19830) \\
    VO (mp-19184) &    MnO (mp-19006) &       HgO (mp-1224) &     HgO (mp-1224) &       YbO (mp-1216) \\
     YN (mp-2114) &     TbN (mp-2117) &        HoN (mp-883) &      HoN (mp-883) &       LuN (mp-1102) \\
    YbO (mp-1216) &     YbS (mp-1820) &       YbS (mp-1820) &     YbS (mp-1820) &       CaO (mp-2605) \\
    YbS (mp-1820) &     YbSe (mp-286) &       YbSe (mp-286) &     YbSe (mp-286) &       CaS (mp-1672) \\
    YbSe (mp-286) &     YbS (mp-1820) &      YbTe (mp-1779) &    YbTe (mp-1779) &      CaSe (mp-1415) \\
   YbTe (mp-1779) &     YbSe (mp-286) &       YbSe (mp-286) &     YbSe (mp-286) &      CaTe (mp-1519) \\
    ZnO (mp-2133) &    ZnSe (mp-1190) &       MgO (mp-1265) &     MgO (mp-1265) &       MgO (mp-1265) \\
   ZnSe (mp-1190) &    CdSe (mp-2691) &      ZnTe (mp-2176) &    ZnTe (mp-2176) &   MgSe (mp-1018040) \\
   ZnTe (mp-2176) &     CdTe (mp-406) &      ZnSe (mp-1190) &    ZnSe (mp-1190) &      MgTe (mp-1039) \\
\end{longtable}

\begin{longtable}{lllll}
\caption{Table of the target formulas from the Materials Project and the template materials used to predict the structure of the target material. The template materials can be considered the most similar under each representation.}
\label{stab2}\\
\toprule
   target\_formula & skipatom\_template &   magpie\_template & megnet16\_template &  hautier\_template \\
\midrule
\endfirsthead
\caption[]{Table of the target formulas from the Materials Project and the template materials used to predict the structure of the target material. The template materials can be considered the most similar under each representation.} \\
\toprule
   target\_formula & skipatom\_template &   magpie\_template & megnet16\_template &  hautier\_template \\
\midrule
\endhead
\midrule
\multicolumn{5}{r}{{Continued on next page}} \\
\midrule
\endfoot

\bottomrule
\endlastfoot
  AgBr (mp-23231) &   CuBr (mp-22913) &    AgI (mp-22925) &  AgO (mp-1079720) &    AgI (mp-22925) \\
  AgCl (mp-22922) &   CuCl (mp-22914) &   CuCl (mp-22914) &  AgO (mp-1079720) &    AgI (mp-22925) \\
   AgI (mp-22925) &   AgCl (mp-22922) &   AgBr (mp-23231) &   AgBr (mp-23231) &   AgBr (mp-23231) \\
 AgO (mp-1079720) &   AgCl (mp-22922) &   AgCl (mp-22922) &   AgCl (mp-22922) &              None \\
   AlAs (mp-2172) &     AlP (mp-1550) &    AlSb (mp-2624) &     AlP (mp-1550) &              None \\
     AlN (mp-661) &     AlP (mp-1550) &     AlP (mp-1550) &     LuN (mp-1102) &      GaN (mp-804) \\
    AlP (mp-1550) &    AlAs (mp-2172) &      BP (mp-1479) &    AlAs (mp-2172) &              None \\
   AlSb (mp-2624) &      AlN (mp-661) &    AlAs (mp-2172) &     AlP (mp-1550) &              None \\
   BAs (mp-10044) &      BP (mp-1479) &    AlAs (mp-2172) &      BP (mp-1479) &              None \\
     BP (mp-1479) &    BAs (mp-10044) &     AlP (mp-1550) &    BAs (mp-10044) &              None \\
    BaO (mp-1342) &     VO (mp-19184) &     SrO (mp-2472) &     SrO (mp-2472) &     BaS (mp-1500) \\
    BaS (mp-1500) &    BaSe (mp-1253) &     SrS (mp-1087) &     SrS (mp-1087) &    BaSe (mp-1253) \\
   BaSe (mp-1253) &     BaS (mp-1500) &    SrSe (mp-2758) &    SrSe (mp-2758) &     BaS (mp-1500) \\
   BaTe (mp-1000) &    SrTe (mp-1958) &    SrTe (mp-1958) &    SrTe (mp-1958) &              None \\
    BeO (mp-2542) &     MgO (mp-1265) &      BeS (mp-422) &      BeS (mp-422) &      BeS (mp-422) \\
     BeS (mp-422) &    BeSe (mp-1541) &     MgS (mp-1315) &     BeO (mp-2542) &    BeSe (mp-1541) \\
   BeSe (mp-1541) &      BeS (mp-422) & MgSe (mp-1018040) &      BeS (mp-422) &      BeS (mp-422) \\
    BeTe (mp-252) &    MgTe (mp-1039) &    MgTe (mp-1039) &    BeSe (mp-1541) &              None \\
    CaO (mp-2605) &     SrO (mp-2472) &     SrO (mp-2472) &     YbO (mp-1216) &     CaS (mp-1672) \\
    CaS (mp-1672) &    CaSe (mp-1415) &     SrS (mp-1087) &     YbS (mp-1820) &    CaSe (mp-1415) \\
   CaSe (mp-1415) &     CaS (mp-1672) &    SrSe (mp-2758) &     YbSe (mp-286) &     CaS (mp-1672) \\
   CaTe (mp-1519) &    SrTe (mp-1958) &    SrTe (mp-1958) &    YbTe (mp-1779) &              None \\
     CdS (mp-672) &    CdSe (mp-2691) &    CdSe (mp-2691) &    CdSe (mp-2691) &    CdSe (mp-2691) \\
   CdSe (mp-2691) & HgSe (mp-1018722) &    ZnSe (mp-1190) &      CdS (mp-672) &      CdS (mp-672) \\
    CdTe (mp-406) &     HgTe (mp-358) &    ZnTe (mp-2176) &    ZnTe (mp-2176) &              None \\
   CoO (mp-22408) &    NiO (mp-19009) &    NiO (mp-19009) &    NiO (mp-19009) &    NiO (mp-19009) \\
   CsAu (mp-2667) &   RbAu (mp-30373) &   RbAu (mp-30373) &   RbAu (mp-30373) &              None \\
 CsBr (mp-571222) &   RbBr (mp-22867) &   RbBr (mp-22867) &   RbBr (mp-22867) &   CsI (mp-614603) \\
 CsCl (mp-573697) &   RbCl (mp-23295) &   RbCl (mp-23295) &     CsF (mp-1784) &   CsI (mp-614603) \\
    CsF (mp-1784) &     RbF (mp-2064) &     RbF (mp-2064) &  CsCl (mp-573697) &  CsCl (mp-573697) \\
 CsH (mp-1057286) &    RbH (mp-24721) &    RbH (mp-24721) &    RbH (mp-24721) &              None \\
  CsI (mp-614603) &    RbI (mp-22903) &    RbI (mp-22903) &    RbI (mp-22903) &    RbI (mp-22903) \\
   CsTe (mp-8361) &              None &              None &              None &              None \\
  CuBr (mp-22913) &   AgBr (mp-23231) &   AgBr (mp-23231) &   AgBr (mp-23231) &   AgBr (mp-23231) \\
  CuCl (mp-22914) &   AgCl (mp-22922) &   AgCl (mp-22922) &   AgCl (mp-22922) &   CuBr (mp-22913) \\
   CuH (mp-24093) &   CuBr (mp-22913) &    LiH (mp-23703) &   CuBr (mp-22913) &              None \\
    DyN (mp-1410) &     TbN (mp-2117) &     TbN (mp-2117) &    ErN (mp-19830) &    ErN (mp-19830) \\
   ErN (mp-19830) &     TmN (mp-1975) &      HoN (mp-883) &     TmN (mp-1975) &     DyN (mp-1410) \\
   GaAs (mp-2534) &     GaP (mp-2490) &     GaP (mp-2490) &     GaP (mp-2490) &              None \\
     GaN (mp-804) &      YN (mp-2114) &     GaP (mp-2490) &    GaAs (mp-2534) &      AlN (mp-661) \\
    GaP (mp-2490) &    GaAs (mp-2534) &    InP (mp-20351) &    GaAs (mp-2534) &              None \\
    GeTe (mp-938) &    SnTe (mp-1883) &    SnTe (mp-1883) &   TePb (mp-19717) &              None \\
    HgO (mp-1224) &      HgS (mp-634) &      HgS (mp-634) &      HgS (mp-634) &      HgS (mp-634) \\
     HgS (mp-634) & HgSe (mp-1018722) &     HgO (mp-1224) &     HgO (mp-1224) & HgSe (mp-1018722) \\
HgSe (mp-1018722) &    CdSe (mp-2691) &     HgTe (mp-358) &    CdSe (mp-2691) &      HgS (mp-634) \\
    HgTe (mp-358) &     CdTe (mp-406) &   TePb (mp-19717) &     CdTe (mp-406) &              None \\
     HoN (mp-883) &     TmN (mp-1975) &     DyN (mp-1410) &     TbN (mp-2117) &     TmN (mp-1975) \\
   InP (mp-20351) &     GaP (mp-2490) &     GaP (mp-2490) &     AlP (mp-1550) &              None \\
   KBr (mp-23251) &   RbBr (mp-22867) &   RbBr (mp-22867) &   RbBr (mp-22867) &     KI (mp-22898) \\
   KCl (mp-23193) &   RbCl (mp-23295) &   RbCl (mp-23295) &   RbCl (mp-23295) &     KI (mp-22898) \\
      KF (mp-463) &     RbF (mp-2064) &     RbF (mp-2064) &     RbF (mp-2064) &    KCl (mp-23193) \\
    KH (mp-24084) &    RbH (mp-24721) &    RbH (mp-24721) &    RbH (mp-24721) &              None \\
    KI (mp-22898) &    RbI (mp-22903) &    RbI (mp-22903) &    RbI (mp-22903) &    KBr (mp-23251) \\
  LiBr (mp-23259) &   NaBr (mp-22916) &   NaBr (mp-22916) &   NaBr (mp-22916) &   LiI (mp-570935) \\
  LiCl (mp-22905) &   NaCl (mp-22862) &   NaCl (mp-22862) &     LiF (mp-1138) &   LiI (mp-570935) \\
    LiF (mp-1138) &      NaF (mp-682) &      NaF (mp-682) &   LiCl (mp-22905) &   LiCl (mp-22905) \\
   LiH (mp-23703) &    NaH (mp-23870) &    NaH (mp-23870) &    NaH (mp-23870) &              None \\
  LiI (mp-570935) &    NaI (mp-23268) &    NaI (mp-23268) &   LiBr (mp-23259) &   LiBr (mp-23259) \\
    LuN (mp-1102) &    ErN (mp-19830) &    ErN (mp-19830) &     TmN (mp-1975) &     TmN (mp-1975) \\
    MgO (mp-1265) &     BeO (mp-2542) &     MgS (mp-1315) &     CaO (mp-2605) &     MgS (mp-1315) \\
    MgS (mp-1315) & MgSe (mp-1018040) &     MgO (mp-1265) &     CaS (mp-1672) & MgSe (mp-1018040) \\
MgSe (mp-1018040) &     MgS (mp-1315) &    MgTe (mp-1039) &    CaSe (mp-1415) &     MgS (mp-1315) \\
   MgTe (mp-1039) &     BeTe (mp-252) & MgSe (mp-1018040) &    CaTe (mp-1519) &              None \\
   MnO (mp-19006) &    NiO (mp-19009) &     VO (mp-19184) &    MnSe (mp-2293) &    MnSe (mp-2293) \\
   MnSe (mp-2293) &     YbSe (mp-286) &    MnO (mp-19006) &    MnO (mp-19006) &    CdSe (mp-2691) \\
  NaBr (mp-22916) &   LiBr (mp-23259) &    KBr (mp-23251) &   LiBr (mp-23259) &    NaI (mp-23268) \\
  NaCl (mp-22862) &   LiCl (mp-22905) &    KCl (mp-23193) &      NaF (mp-682) &    NaI (mp-23268) \\
     NaF (mp-682) &     LiF (mp-1138) &       KF (mp-463) &   NaCl (mp-22862) &   NaCl (mp-22862) \\
   NaH (mp-23870) &    LiH (mp-23703) &     KH (mp-24084) &    LiH (mp-23703) &              None \\
   NaI (mp-23268) &   LiI (mp-570935) &     KI (mp-22898) &   LiI (mp-570935) &   NaBr (mp-22916) \\
   NiO (mp-19009) &    CoO (mp-22408) &    CoO (mp-22408) &    CoO (mp-22408) &     VO (mp-19184) \\
   PbS (mp-21276) &    PbSe (mp-2201) &      HgS (mp-634) &    PbSe (mp-2201) &    PbSe (mp-2201) \\
   PbSe (mp-2201) &    PbS (mp-21276) & HgSe (mp-1018722) &    PbS (mp-21276) &    PbS (mp-21276) \\
  RbAu (mp-30373) &    CsAu (mp-2667) &    CsAu (mp-2667) &    CsAu (mp-2667) &              None \\
  RbBr (mp-22867) &    KBr (mp-23251) &  CsBr (mp-571222) &    KBr (mp-23251) &    RbI (mp-22903) \\
  RbCl (mp-23295) &    KCl (mp-23193) &  CsCl (mp-573697) &    KCl (mp-23193) &    RbI (mp-22903) \\
    RbF (mp-2064) &       KF (mp-463) &     CsF (mp-1784) &       KF (mp-463) &   RbCl (mp-23295) \\
   RbH (mp-24721) &     KH (mp-24084) &  CsH (mp-1057286) &     KH (mp-24084) &              None \\
   RbI (mp-22903) &     KI (mp-22898) &   CsI (mp-614603) &     KI (mp-22898) &   RbBr (mp-22867) \\
   SnTe (mp-1883) &     GeTe (mp-938) &     GeTe (mp-938) &    ZnTe (mp-2176) &              None \\
    SrO (mp-2472) &     CaO (mp-2605) &     BaO (mp-1342) &     CaO (mp-2605) &     SrS (mp-1087) \\
    SrS (mp-1087) &    SrSe (mp-2758) &     BaS (mp-1500) &     CaS (mp-1672) &    SrSe (mp-2758) \\
   SrSe (mp-2758) &     SrS (mp-1087) &    BaSe (mp-1253) &    CaSe (mp-1415) &     SrS (mp-1087) \\
   SrTe (mp-1958) &    CaTe (mp-1519) &    BaTe (mp-1000) &    CaTe (mp-1519) &              None \\
    TbN (mp-2117) &     DyN (mp-1410) &     DyN (mp-1410) &     DyN (mp-1410) &     TmN (mp-1975) \\
  TePb (mp-19717) &    SnTe (mp-1883) &     HgTe (mp-358) &     CdTe (mp-406) &              None \\
 TlBr (mp-568560) &  TlCl (mp-569639) &   TlI (mp-571102) &   TlI (mp-571102) &   TlI (mp-571102) \\
 TlCl (mp-569639) &   TlI (mp-571102) &      TlF (mp-720) &      TlF (mp-720) &   TlI (mp-571102) \\
     TlF (mp-720) &       KF (mp-463) &  TlCl (mp-569639) &  TlCl (mp-569639) &  TlCl (mp-569639) \\
  TlI (mp-571102) &  TlCl (mp-569639) &  TlBr (mp-568560) &  TlBr (mp-568560) &  TlBr (mp-568560) \\
    TmN (mp-1975) &    ErN (mp-19830) &    ErN (mp-19830) &    ErN (mp-19830) &     TbN (mp-2117) \\
    VO (mp-19184) &     BaO (mp-1342) &    MnO (mp-19006) &    MnO (mp-19006) &    NiO (mp-19009) \\
     YN (mp-2114) &      GaN (mp-804) &     LuN (mp-1102) &     DyN (mp-1410) &     TbN (mp-2117) \\
    YbO (mp-1216) &     YbS (mp-1820) &     YbS (mp-1820) &     CaO (mp-2605) &     YbS (mp-1820) \\
    YbS (mp-1820) &     YbSe (mp-286) &     YbO (mp-1216) &     CaS (mp-1672) &     YbSe (mp-286) \\
    YbSe (mp-286) &     YbS (mp-1820) &    YbTe (mp-1779) &    CaSe (mp-1415) &     YbS (mp-1820) \\
   YbTe (mp-1779) &    ZnTe (mp-2176) &     YbSe (mp-286) &    CaTe (mp-1519) &              None \\
    ZnO (mp-2133) &     YbO (mp-1216) &    ZnSe (mp-1190) &    ZnSe (mp-1190) &    ZnSe (mp-1190) \\
   ZnSe (mp-1190) &    CdSe (mp-2691) &    CdSe (mp-2691) &     ZnO (mp-2133) &    CdSe (mp-2691) \\
   ZnTe (mp-2176) &     CdTe (mp-406) &     CdTe (mp-406) &     CdTe (mp-406) &              None \\
\end{longtable}